\documentclass[%
 aip,
 amsmath,amssymb,
 reprint,%
]{revtex4-1}

\usepackage{graphicx}
\usepackage{dcolumn}
\usepackage{bm}
\usepackage[utf8]{inputenc}
\usepackage[T1]{fontenc}
\usepackage{mathptmx}
\usepackage{etoolbox}
\usepackage{amsmath}
\usepackage[usenames,dvipsnames]{color}

\usepackage{physics}
\usepackage{mathtools}
\usepackage{dsfont}
\makeatletter
\def\@email#1#2{%
 \endgroup
 \patchcmd{\titleblock@produce}
  {\frontmatter@RRAPformat}
  {\frontmatter@RRAPformat{\produce@RRAP{*#1\href{mailto:#2}{#2}}}\frontmatter@RRAPformat}
  {}{}
}%
\makeatother
\begin{document}

\preprint{AIP/123-QED}

\title[Emulating quantum computing with optical matrix multiplication]{Emulating quantum computing with optical matrix multiplication}
\author{Mwezi Koni}
\affiliation{School of Physics, University of the Witwatersrand, Private Bag 3, Wits 2050, South Africa}

\author{Hadrian Bezuidenhout}
\affiliation{School of Physics, University of the Witwatersrand, Private Bag 3, Wits 2050, South Africa}

\author{Isaac Nape}
\affiliation{School of Physics, University of the Witwatersrand, Private Bag 3, Wits 2050, South Africa}
\email{isaac.nape@wits.ac.za}

\newcommand{\IN}[1]{\textcolor{blue}{ #1}}
\newcommand{\note}[1]{\textcolor{blue}{ #1}}

\date{\today}

\begin{abstract}
Optical computing harnesses the speed of light to perform vector-matrix operations efficiently. It leverages interference, a cornerstone of quantum computing algorithms, to enable parallel computations.  In this work, we interweave quantum computing with classical structured light  by formulating the process of photonic matrix multiplication using quantum mechanical principles such as state superposition and subsequently demonstrate a well known algorithm, namely the Deutsch-Jozsa's algorithm. This is accomplished by elucidating the inherent tensor product structure within the Cartesian transverse degrees of freedom of light, which is the main resource for optical vector-matrix multiplication.   To this end, we establish a discrete basis using localized Gaussian modes arranged in a lattice formation and demonstrate the operation of a Hadamard Gate. Leveraging the reprogrammable and digital capabilities of spatial light modulators, coupled with Fourier transforms by lenses, our approach proves adaptable to various algorithms. Therefore, our work advances the use of structured light for quantum information processing.
\end{abstract}

\maketitle

\section{Introduction}

Controlling various degrees of freedom of light, i.e. time, frequency, spatial and  momentum, has  become an emerging and promising tool for numerous information processing tasks in classical and quantum domains, ranging from novel imaging methods \cite{cameron2024adaptive, moodley2024advances, geng2011structured, erkmen2010ghost, ferri2010differential, zia2023interferometric, kim2022metasurface}, communications \cite{mirhosseini2015high, wang2012terabit, krenn2014communication, wang2022orbital, forbes2024quantum, lavery2017free} and computation \cite{perez2018first, peruzzo2014variational, konno2024logical, mcmahon2023physics}, all harnessing the high dimensional nature  of structured light fields \cite{nape2023quantum, forbes2021structured, rubinsztein2016roadmap, piccardo2021roadmap, bliokh2023roadmap}. This is because spatial modes enable for information encoding in qudit  spaces for dimensions $d>2$ per particle as opposed to the $d=2$ (qubits) encoding levels offered by traditional qubit encoding such as with polarisation states. Because qubits are easy to control, they are a reliable resource for numerous protocols.

In traditional  quantum computing, encoding basis states are typically formed from multiple qubits ($d=2$, states) and the operations needed to construct arbitrary unitary gates in $d>2$  dimensions often involve a mix of  two-dimensional gate operations (such as the Hadamard gate and T-gate) along with a multi-qubit operations (like the Control-not gate) \cite{boykin2000new}.  This  allows for precise and efficient execution of arbitrary computations that are required to demonstrate quantum advantage  \cite{kitaev2002classical}.   Given that each qubit is restricted to two dimensions, the only way to enhance encoding capacity is by increasing the number of qubits and ensuring optimal connectivity among them. An alternative is to utilize qudits, which are particles that occupy $d>2$ dimensions, because they offer larger encoding state spaces \cite{wang2020qudits} and promise robustness against noise \cite{ecker2019overcoming}. As a result, structured photon states  emerge as a promising candidate for achieving this, as they occupy higher-dimensional Hilbert spaces.Furthermore, it has been demonstrated that vector-matrix multiplication, a fundamental operation in quantum mechanics, can be performed using transverse optical fields \cite{spall2020fully}.

There have been numerous impressive proposals and implementations of quantum computing algorithms that utilize the higher dimensional nature of the internal degrees of freedom of photons \cite{ garcia2012quantum, o2007optical, gao2019arbitrary}. This field has developed tremendously, aided by the advent of multiplane light conversion technology \cite{fontaine2022photonic}, where digital holography and free-space diffraction are employed to realize arbitrary unitary gate operations \cite{brandt2020high}. Inspired by these advances, diffraction-based deep neural network architectures have also been proposed \cite{wang2024ultrahigh}. Additionally, harnessing mode mixing in complex media such as multimodal optical fibers, as opposed to diffractive MPLCs, has emerged as a potential avenue (see review on this topic \cite{lib2022quantum}) to achieve similar operations but with discrete pixel-like states \cite{goel2024inverse}. To realise the full potential  of  qudits for quantum computing, efficient single photon sources and the tools for entangling and controlling them are necessary.

On the other hand, harnessing  classical optical fields for quantum computing has also been an active area of research\cite{cerf1998optical, spreeuw2001classical, londero2004efficient, kaur2007optical, perez2016quantum, perez2018first}. The drive in this field has been inspired by the observation that some of the essential resources of quantum computing can be found in classical coherent fields. Like quantum states, classical waves can be prepared in superpositions and can be interfered, allowing for parallel information processing. Notably, Jozsa
 also argued that classical waves can provide an efficient simulation platform since not all quantum algorithms require entanglement as a resource   \cite{jozsa1997entanglement,jozsa2003role}. 
Recent advances have demonstrated that optical methods can implement unitary and nonunitary transformations efficiently, utilising programmable holographic techniques and spatial phase modulation to enable high-dimensional operations \cite{wang2017programmable}. Similarly, parallel processing for computational tasks like multiplication modulo, a crucial component of Shor's algorithm, using phase modulation has been explored, offering promising avenues for structured light applications in computational tasks \cite{nitta2008parallel}. This opens up new possibilities for studying or implementing quantum computational tasks without relying on complex quantum hardware, by merely exploiting coherent fields and performing the required matrix-vector operations on them.  

An intriguing and unexplored approach to utilizing coherent fields in quantum computing is Tamura's optical matrix-vector multiplication from the 1970s \cite{tamura1979two} , which was long forgotten but has recently resurfaced. This method harnesses point-wise multiplication (Hadamard product) of light and the Fourier transform capabilities of lenses. With this approach, operations have been applied in dimensions reaching up to \(d = 56\) dimensions \cite{spall2020fully}. Hidden in this method, is the capability to encode information in Hilbert spaces formed from transverse modes of coherent laser fields.

In this work, we use optical matrix multiplication  as a foundation for emulating quantum algorithms, therefore interacting quantum computing with structured light. we unveil the inherent tensor product structure of transverse spatial modes that is exploited in coherent optical matrix-vector multiplication and show that  it can be harnessed  as a tool to emulate quantum computing algorithms. Our logical basis is constructed from lattices of displaced Gaussian modes, eigenmodes of free space, that are locally modulated with digital holograms encoded on spatial light modulators. We treat the Cartesian location coordinates \(x\) and \(y\) of each mode as our degrees of freedom, forming the equivalent of two qudit registers. The \(x\) component is used to embed our state vectors, while the \(y\)-components acts as an ancilla that facilities the unitary gate operator implementation. After a cylindrical lens integrates over the \(x\)-component, the final resulting state is measured in the \(y\)-coordinate. We test this using our basis and the Hadamard transform as example and demonstrate the Deutsch-Jozsa algorithm with our encoding scheme, showing it can  be used to query balanced and constant function, with average fidelity above $90 \%$. In our scheme, we exploit the inherent parallelism offered by optical matrix-vector multiplication, allowing to prepare multiple identical spatial modes  as a superposition, and encode the oracle matrix as local phase transformations on each state.

\begin{figure*}[t!]
    \centering
\includegraphics[width=0.95\linewidth]{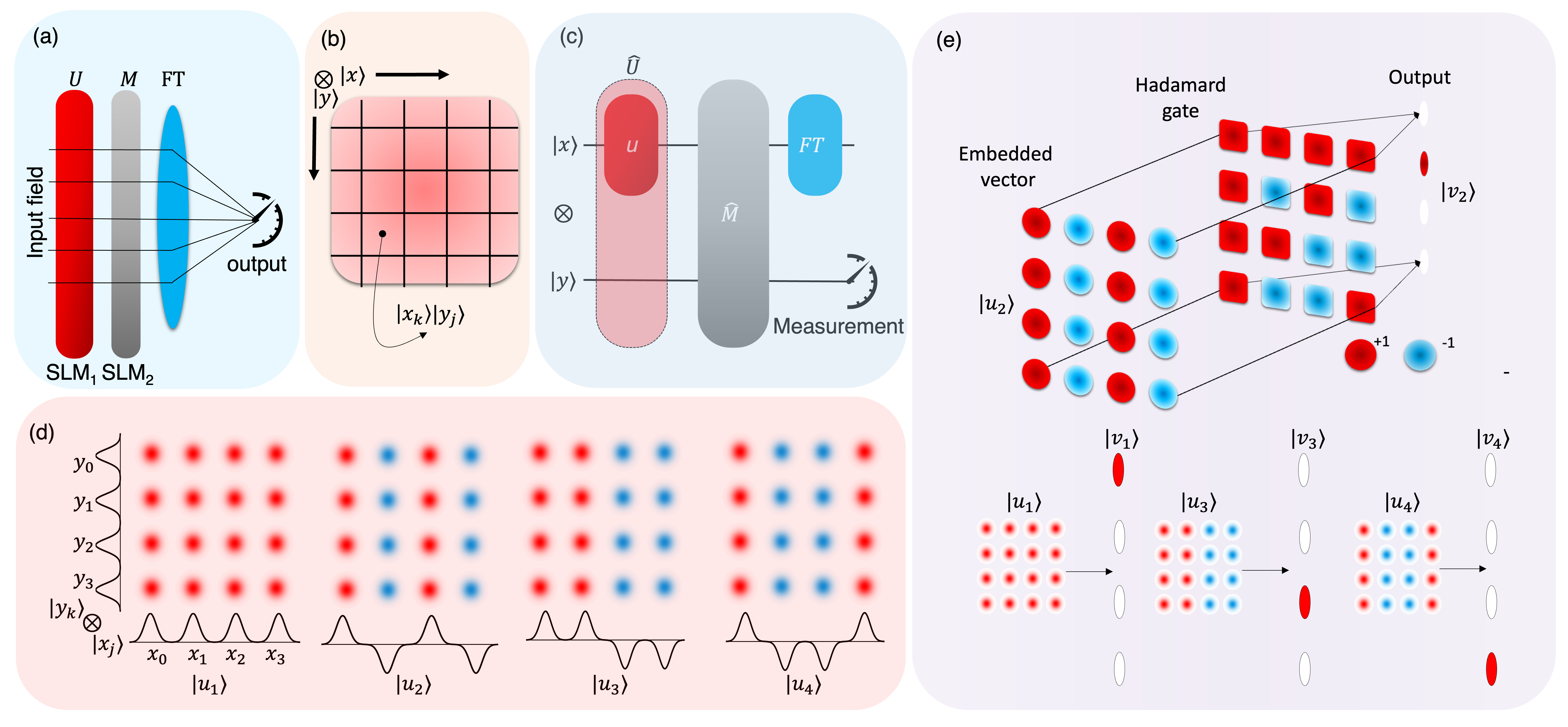}
    \caption{(a) Analogue of quantum computing  using optical vector-matrix multiplication.  $\text{SLM}_1$ encodes the lattice state via an operation, $U$, by copying the vector components encoded in the x-direction into the y-direction following Ref. \cite{spall2020fully}. The second SLM encodes the matrix operation $M$. After this a cylindrical lens focuses the field in x-direction to produce the final matrix-vector product at the output farfield plane. (b) The encodings performed by the spatial light modulators (SLMs) operate on  the tensor product states $\ket{x_k}\otimes \ket{y_j}$ corresponding to the discrete position states that can be obtained by partitioning  the Cartesian plane ($\mathbb{R}_2$) into non-overlapping segments. (c) A quantum circuit for performing matrix multiplication using our basis states $\ket{x_k}\otimes\ket{y_j}$, corresponding to the coordinates of the partitions. The operations are represented as operators that act on the x-coordinate and y-coordinate degrees of freedom. (d) An example of the Hadamard basis states, $\ket{u_j}$, encoded onto the lattice states for N=4. Here, the lattice is formed from propagation invariant Gaussian modes, each encoded into the independent partitions.  To demonstrate that they form a tensor product space, $\mathcal{H}_N \otimes \mathcal{H}_N$,  we show the individual states for the x and y coordinates, where the final lattice is obtained from their tensor products.  These states are orthorgonal.  (e) Example of one of the states being acted on by the Hadamard gate and then focused to perform the column wise summation. The embedded state vector is modulated, pointwise, with an SLM encoding the Hadamard matrix/gate. This is then focused in the farfield with a cylindrical lens where the zero order component of the pattern contains the result in the vertical direction.  The outcomes are shown for other states in the Hadamard basis mapping from $\ket{u_j}$ to the $\ket{v_j}$.}
    \label{fig:concept}
\end{figure*}
\section{Theory}
\subsection{Matrix-vector multiplication via pointwise multiplication}

Consider a matrix, \(M\), with elements \(M_{jk}\), and a vector \(u\) with elements \(u_k\). We can compute the matrix-vector product to produce the vector $v \equiv M u$, where the $j^\text{th}$ component can be computed from $v_j = \sum_k M_{jk}u_k$. The goal is to perform such an operation using light fields. While this technique has already been introduced in classical optics \cite{spall2020fully}, here it will be reformulated for quantum computing, where the vector \(v\) now encodes a normalized quantum state and \(M\) encodes a unitary gate operator. Firstly, notice that the matrix-vector product can be rewritten as:
\begin{equation}
v_j = \sum_k M_{jk} u_k  = \sum_k (M \odot U)_{jk},
\label{eq:prod}
\end{equation}
representing the summation over the columns of the element-wise product (or Hadamard product \cite{horn1990hadamard}), denoted by the symbol $\odot$, between our matrix $M$ and another matrix $U$, which encodes or copies $u$ into its rows. The matrix $U$ can be obtained from the outer product, $U = u^\intercal \otimes \textbf{1}$, with $\textbf{1} = (1,1, 1, 1, \ldots)^\intercal$.

We see that the matrix-vector multiplication can be decomposed into the elementwise product $M \odot U$ and a subsequent summation of the columns. These operations can be achieved optically using spatial light modulators (SLM) and a cylindrical lens (CL) as shown in Fig. \ref{fig:concept}(a) as demonstrated in \cite{spall2020fully}. On one SLM, a uniform field can be modulated with a hologram that encodes (point-wise) the matrix $U$ and is imaged onto a second SLM, which is encoded with the matrix $M$. The imaging system between the two SLMs enables the elementwise multiplication of the encoded matrices. The final operation is performed using a cylindrical lens that focuses the field in the direction that coincides with the columns of $M$, performing a column-wise summation—therefore completing the product to produce the elements $v_j$

Remarkably, hidden in this approach is the fact that the transverse Cartesian coordinates are used as independent degrees of freedom that can be used to enact operations, analogous to non-interacting multi-particle systems in quantum computing. Next, we show how this idea can be used to emulate quantum computing with transverse fields.

\subsection{Formulation using quantum states}

In our approach, we emulate quantum computing by formulating  optical vector-matrix  multiplication in the language of quantum mechanics by representing the optical fields that encode the vectors as states on Hilbert space  and operations acting on the  fields to represent the matrices that enact the equivalent of gate operations as shown in  Fig.  \ref{fig:concept} (b)  and (c).  Firstly, the x-y coordinate system can be decomposed into a tensor-product space using the continuous degrees of freedom mapping the states $\ket{x}\in\mathcal{H}_{\infty}$ and $\ket{y}\in\mathcal{H}_\infty$  so that any field $\psi(x, y)$ is described by the  state $\ket{\psi} = \int \psi(x, y)  \ket{x}\ket{y} dxdy$ where the basis states satisfy  $\braket{x}{x'} = \delta(x-x')$ and $\braket{y}{y'} = \delta(y-y')$. This means that each photon in a laser field is defined on a Hilbert space $\mathcal{H}_\infty \otimes \mathcal{H}_{\infty}$ spanned by the Cartesian coordinate internal degrees of freedom (i.e. the coordinate states).  By partitioning the transverse plane into $N$ bounded and non-overlap intervals , $\ket{x}\rightarrow\ket{x_j}$ and $\ket{y}\rightarrow\ket{y_k}$, shown in Fig. \ref{fig:concept} (b), we can now define qudit states that form a high dimensional qudit space on the combined Hilbert space, i.e. $\mathcal{H}_N \otimes \mathcal{H}_N$ that has dimensions $N^2$.This has been discussed in recent work\cite{shen2022nonseparable, aiello2015quantum}, showing that non-overlapping regions of transverse optical fields can be expressed using a basis formed from tensor product states, much like distinct particles that occupy their own Hilbert spaces. This ensures that the spatial modes corresponding to the $x$ and $y$ coordinates function as independent degrees of freedom.   From these states, a uniform superposition  can be prepared as
\begin{equation}
\ket{\Psi}  = \frac{1}{N} \sum_{jk} \ket{x_k} \otimes\ket{y_j},
  \label{eq:lattice0}
\end{equation}
where each state in the partition has a non-zero coefficient.
From this description, we see that the operations that can be applied on them can be enacted on each individual degree of freedom or on both simultaneously. Using our matrix vector product from Eq.~(\ref{eq:prod}), we can show that using such states and treating the x and y degrees of freedom as analogues of qudit registers,  that we can execute matrix vector (equivalently, operator and state) products. For instance, we can map the matrix $U$ that was defined earlier onto an operator  that encodes information into the lattice as,
\begin{equation} 
\hat{U}  = \sum_{m} u_m \ket{x_m}\bra{x_m} \otimes \mathbb{I},
\label{eq:CoeffEnc}
\end{equation}
this is the analogue for the matrix $\hat{U}$ in Eq.~ (\ref{eq:prod}) -  assuming that the initial state is the superposition $\ket{\Psi}$. This approach leverages multiple channels (Gaussian arrays) to carry independent streams of information simultaneously, reminescent of multiplexing approaches \cite{he2022towards}. Here, the coefficients mark the $x$-components but copy all the elements, $u_m$, across the $y$-components - this still leaves the $y$-components independent of the $x$-components. Similarly,  for matrix $M$ we have the operator 
\begin{equation}
   \hat{M} = \sum_{mn} M_{nm} \ket{x_m}\ket{y_n} \bra{y_n}\bra{x_m},
   \label{eq:operator}
\end{equation}
which instead interacts with both components (registers),  therefore encoding the matrix component's $M_{nm}$  into each state $\ket{x_m}\ket{y_n}$ of the lattice. The combined operation $\hat{M}\hat{U}$ produces the state, 
\begin{equation}
\ket{\Psi}  = \frac{1}{N} \sum_{jk} M_{jk} u_k  \ket{x_k} \otimes\ket{y_j}.
  \label{eq:lattice2}
\end{equation}
which has the same components as Eq.~(\ref{eq:prod})  $v_j =\sum_k M_{jk} u_k$  from before. While the result of the multiplication is encoded into the lattice, which is $N \times N$ in dimensions, the result can be read out as a column in the y-direction once the summation $k$ (in the x-component) is contracted. This can be done using the Fourier transforming (FT) lens in the x-direction - the analogue of a quantum Fourier transform, therefore completing the circuit in Fig.~\ref{fig:concept} (c). The result of this is an interference pattern, having a zero order component that corresponds to the  amplitude,  $\sum_{jk} e^{i f_{x_l}x_k} M_{jk} v_k$ for the frequency component $f_{x_l} = 0$ . Measuring the zeroth order interference pattern in x-direction integrates the x-component, therefore leaving the y-components in the state
\begin{equation}
\ket{v}  = \frac{1}{N} \sum_{jk} M_{jk} u_k \ket{y_j}.
  \label{eq:lattice3}
\end{equation}
\begin{figure*}[ht!]
    \centering
\includegraphics[width=0.95\linewidth]{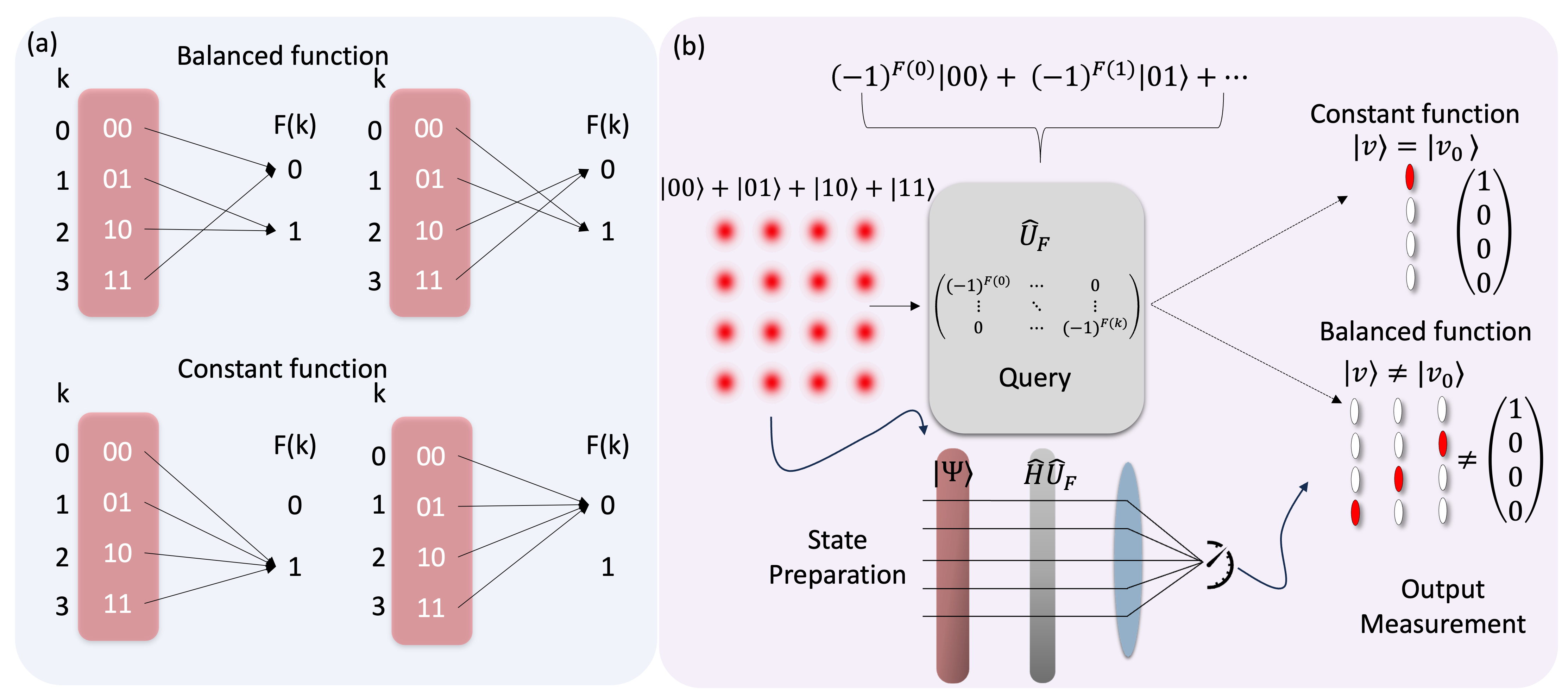}
    \caption{(a) Conceptual illustration of the  balanced and constant functions involved in the Deutsch-Jozsa Algorithm for binary strings of length two, i.e.  00, 01, 10, 11. The two top figures showcase a balanced function which maps half of its inputs to $0$ and the other half to $1$ and the two bottom figure, a constant function that maps all its inputs to either $0$ or  to $1$.  (b) Illustration of our optical implementation of the Deutsch-Jozsa Algorithm. A lattice of Gaussian beams, representing the superposition of states, is simultaneously prepared by Alice. The state is then queried by Bob in parallel by the action of the compound oracle and hadarmard operations, which impart function evaluations spatially on the state through phase transformations and map us back into the computational basis. A constant function returns the state mapping onto the binary string $00$, indicated by the vector lighting up on the first elements. A balanced function returns a state orthogonal to $00$, shown by light lobes at different positions. }
    \label{fig:Joza_concept}
\end{figure*}

as expected. Next, measuring each y-components produces the probability amplitudes  $v_j = \sum_k M_{jk} u_k$, having the same outcome as  $v = Mu$ from Eq.~(\ref{eq:prod}).  Because this scheme allows us to express operations of the x-y coordinates analogous to gate based computing platforms, we will exploit it for quantum computing, where the matrix mechanics that describes quantum computing can be formulated using the transverse spatial modes in Cartesian coordinates. Next we introduce our basis of choice for doing this and illustrate the Hadamard gate.

\section{Hadamard Gate}
\label{sec: hadamard}

Before we introduce the Hadamard gate we first present our encoding basis. Instead of using arrays of square pixel like states for each bounded interval on the x-y plane, we encode arrays of Gaussian modes, as shown in Fig.~ \ref{fig:concept}(c). The illustrations show that our Gaussian lattices/arrays are encoded as tensor products of arrays of one dimensional Gaussian modes, where  $(x_j, y_k) \rightarrow \ket{x_j}\ket{y_k}$ denote the centres of each Gaussian mode given by the states,
\begin{align}
  \ket{x_j }\ket{y_k} &\propto  \int_{A_j}  e^{-\frac{(x-x_j)^2}{w^2}} \ket{x} dx   \otimes \int_{A_k} e^{-\frac{(y-y_k)^2}{w^2}}  \ket{y}dy \nonumber \\ 
& =  \int_{A_j \times A_k}  e^{-\frac{(x-x_j)^2 + (y-y_k)^2}{w^2}}  \ket{x}\ket{y} dxdy,
\end{align}
where each mode is centred at coordinates $(x_j, y_k)$  within a closed  boundaries $A_j \times A_k \subset \mathbb{R}_2$ that are non overlapping. In this way, uniform superpositions such as in Eq. ~ (\ref{eq:lattice0}) can  also be prepared, i.e. resulting in the field in the first panel of Fig. \ref{fig:concept} (d). The rest of the fields in the figure represent instance of the Hadamard basis states, which include the uniform superposition. Next, we show how the Hadamard gate can be enacted on these fields.

The N-dimensional Hadamard gate, $\hat{H}_N$, is commonly used for quantum computing to map between the logical and the superposition basis (or vice-versa). For example, for two-dimensional states 
 ($N=2$),  we can describe the logical basis as  $\ket{v_0} \equiv \ket{0}= (1, 0)^{\intercal}$ and $\ket{v_1} \equiv\ket{1} = (0, 1)^{\intercal}$,. In this basis the Hadamard gate is given by, $\hat{H}_2 = \frac{1}{\sqrt{2}} \begin{pmatrix}1 & 1\\ 1 & -1\end{pmatrix}$. Applying the Hadamard gate maps the logical basis states onto the Hadamard basis  states $\ket{u_{1,2}} = 1/\sqrt{2}(\ket{0}\pm \ket{1})$, respectively. The states $\ket{u_{1,2}}$ are orthogonal and form a  basis. The Hadamard gate performs a change of basis, however when applied twice it leaves the state unchanged, i.e $\hat{H}\hat{H} = \mathbb{I}$, true for all dimensions N.   For qudits, $N>2$, the Hadamard matrix can be computed for dimensions $N =2^n$ ($n$ power of 2) as $\hat{H}_{N} = \hat{H}^{\otimes n}$, using tensor products of $n$ Hadamard gates. Thus, the $k^{th}$ Hadamard basis state can be obtained by applying the Hadamard gate to the $k^{th}$ state in the logical basis following, 
 \begin{align}
      \hat{H}_N \ket{v_k} &=  \ket{u_k} \nonumber \\
          &= \frac{1}{\sqrt{N}} \sum^{N-1}_{m =0} (-1)^{m \cdot k }  \ket{m},
 \end{align}
where the states $\ket{m}$ are the standard basis states with coefficients $(-1)^{m \cdot k}$ with  $m \cdot k$ corresponding  to the dot product of the binary representation of  the integers $m$ and $k$.  Accordingly, the Hadamard basis states can be mapped onto our lattice states as
\begin{equation}
\ket{u_k} =  \frac{1}{N  }\sum^{N-1}_{mn = 0}  (-1)^{m \cdot k}\ket{x_m}\ket{y_n}.
    \label{eq:hadamardstate}
\end{equation}
Figure~\ref{fig:concept}~ (d) shows examples of the lattice states that have been encoded with the Hadamard basis states in $N=4$ dimensions. 
To encode these coefficients into the lattice, we use the operator $\hat{U}$ in Eq. \ref{eq:CoeffEnc}, but with the coefficients corresponding to our Hadamard basis states. Applying the Hadamard  gate to each of the states in the  lattice should map  them back to the logical basis states, following $\hat{H}_N\ket{u_k}=\ket{v_k}$.  Each basis state $\ket{u_k}$ that is embedded in the lattice (as in Fig. \ref{fig:concept} (d)) can  be mapped onto a unique element vector with one non-zero entry,  as shown in Fig. \ref{fig:concept} (e). To encode the Hadamard gate, we can map it's matrix components  onto the lattice, using Eq.~(\ref{eq:operator})  where  $M_{jk}$ will represent the matrix elements of the Hadamard gate. Thereafter, we apply the Fourier transform and the zero order component filtering to complete the operator-state (matrix-vector) multiplication.

Next we show how this scheme can be used to perform a well known quantum algorithm.

\subsection{Application: Deutsch-Jozsa algorithm}

 The Deutsch-Jozsa algorithm \cite{deutsch1992rapid} provides an efficient quantum solution for determining whether a boolean function, $f(\cdot)$, that takes binary values as inputs, is balanced (half of the inputs map to 1 and the other half to 0) or constant (all inputs either map to 0 or 1). 
To illustrate this, consider a collection of 2-bit input strings chosen from the set $\mathcal{B} = \{00, 01, 10, 11\}$; each element $k \in \mathcal{B}$   can  be used an input to the function, i.e. $F(k)$.  A balanced function would return 0 for half the entries, e.g., $k \in\{00, 11\}$, and 1 for the other half, $k \in \{01, 10\}$, as shown in the first left panel of Fig. \ref{fig:Joza_concept} (a). The second example (top right panel of Fig. \ref{fig:Joza_concept} (a)) shows an alternative combination that can also represent a balanced function. For the constant case, all binary values map to a single output, i.e., $\{00, 01, 10, 11\} \rightarrow 0$ or $\{00, 01, 10, 11\} \rightarrow 1$ as shown in the bottom panel of Fig.~\ref{fig:Joza_concept} (b) for two instances of a such a function.  

To query the nature of such a function, the quantum computer prepares a superposition of all possible binary values in $\mathcal{B}$, queries the function values simultaneously, and through interference it returns a result indicating whether the function is balanced or not. We adapt the algorithm to our encoding scheme using the lattice basis and our operator-state (matrix-vector) multiplication scheme. 

\begin{figure*}[ht!]
    \centering
\includegraphics[width=0.95\linewidth]{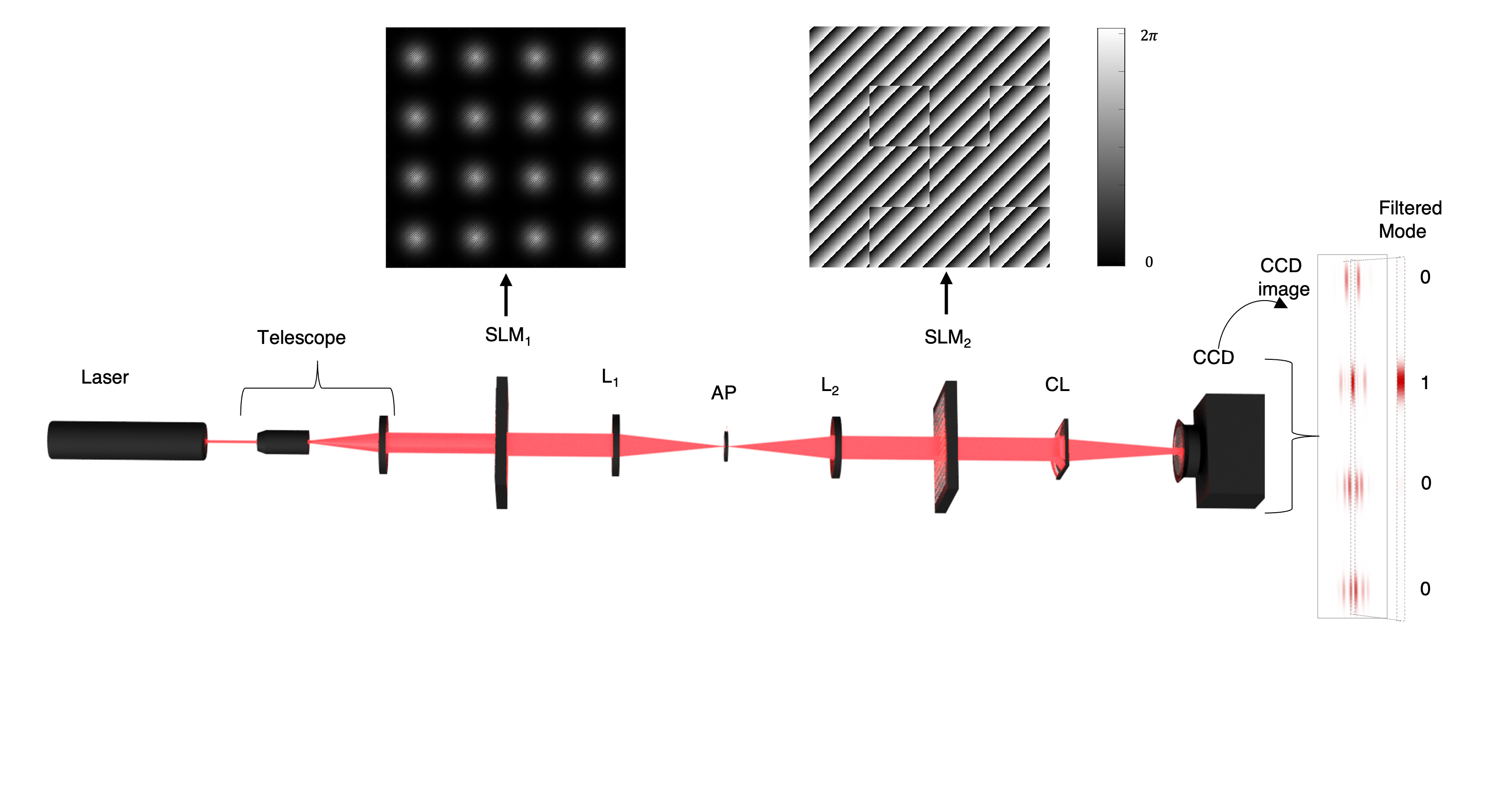}
    \caption{A laser beam is expanded using a telescope to overfill the active area of the first Spatial Light Modulator (SLM 1). At this point, the beam is transformed to encode the input vector as repeated rows in a matrix of Gaussian beams. This output is then relayed to SLM 2 using the 4f system of lenses (L$_2$ and L$_1$), where the lattice undergoes an additional phase encoding. An aperture (AP) at the focal plane of lens L$_1$ filters out the first-order diffraction pattern. Subsequently, the beam is propagated through a cylindrical lens (CL) and undergoes a one-dimensional Fourier transform in the x-coordinate. The final outcome is captured by a CCD camera, with the result of the multiplication found on the central fringe, which is filtered out (as shown in the inset). The filtered mode image on the CCD camera displays the result of the matrix-vector multiplication, where regions of no intensity correspond to zero elements, and bright regions correspond to non-zero elements. The holograms encoded on each SLM are shown as grayscale images.}
    \label{fig:setup}
\end{figure*}

We illustrate the concept using four-dimensional  (N=4) states as an example. We can map the binary values of the inputs to the function $F(\cdot)$ in the set $\mathcal{B}$ to the logical basis states

\begin{align}
\ket{0} \rightarrow  \ket{00} &= 
\begin{pmatrix}
1 \\ 0 \\ 0 \\ 0 
\end{pmatrix}, 
\ket{1} \rightarrow \ket{01} = 
\begin{pmatrix}
0 \\ 1 \\ 0 \\ 0 
\end{pmatrix}, \nonumber\\
\ket{2} \rightarrow \ket{10} &= 
\begin{pmatrix}
0 \\ 0 \\ 1 \\ 0 
\end{pmatrix},
\ket{3} \rightarrow \ket{11} = 
\begin{pmatrix}
0 \\ 0 \\ 0 \\ 1 
\end{pmatrix}.
\end{align}

Using this basis, we can prepare all the possible inputs as the un-normalised superposition state $\ket{0} + \ket{1} + \ket{2} + \ket{3}$ where the binary values are mapped onto their integer representations. These correspond to the logical basis states $\ket{v_k}$ from the previous section. In the algorithm, we will see that the mappings of the gates convert between the Hadamard basis states ($\ket{u_k}$) and the logical basis states  ($\ket{v_k}$). To prepare the uniform superposition ($\ket{0} + \ket{1} + \ket{2} + \ket{3}$) using our lattice, we can use the first element of the Hadamard basis following  Eq.~(\ref{eq:hadamardstate})
\begin{equation}
\ket{u_0}  = \frac{1}{N}  \sum^{N-1}_{jk = 0} \ket{x_k} \otimes\ket{y_j}.
  \label{eq:lattice200}
\end{equation}
This indeed encodes the uniform superposition into our lattice as shown in Fig.~\ref{fig:Joza_concept} (b).
Next, the features of the function $F(k)$, are encoded onto the unitary operator as $U_F$ so that  the states in the superposition become $\sum_k  (-1)^{F(k)} \ket{k}$, showing that $U_F$ is diagonal, and has the form
\begin{equation}
  U_F = \begin{pmatrix}
    (-1)^{F(0)} & 0 & \dots & 0 \\
    0 & (-1)^{F(1)} & \dots & 0 \\
    \vdots & \vdots & \ddots & \vdots \\
    0 & 0 & \dots & (-1)^{F(N-1)}
  \end{pmatrix},
\end{equation}
where each state in the superposition  is marked with a coefficient $(-1)^{F(k)}$.  As such, we see that $U_F$ encodes a $\pi$ phases depending on whether $F(k)$ is 0 or 1.   Motivated by this, we encode the unitary as $\hat{U}_F = \sum_{k} (U_F)_{kk} \ket{x_k}\bra{x_k} \otimes \mathbb{I}$ onto our lattice, where $(U_F)_{kk}$  are the diagonal elements of $U_F$ ,  marking the x-components of the lattice with the desired phases. In this way we do not need ancillary particles thanks to the higher dimensional nature of our degrees of freedom.

Accordingly,  given the uniform superposition of the lattice, we obtain 
\begin{align}
\hat{U}_F \ket{u_0} &= \frac{1}{N} ( (-1)^{F(0)} \ket{x_0}\ket{y_0} + (-1)^{F(1)}  \ket{x_1}\ket{y_0} + ..) \\
&= \frac{1}{N}\sum^{N-1}_{jk = 0} (-1)^{F(k)} \ket{x_k}\ket{y_j}.
\end{align}
As a result, we can express the state as a piece wise function
\begin{equation}
\hat{U}_F \ket{u_0} =
\begin{cases}
\pm \ket{u_0}, & \text{Constant} \\
 \frac{1}{N} \sum_{jk} (-1)^{F(k)} \ket{x_k}\ket{y_j} & \text{Balanced},
\end{cases}
\end{equation}
\noindent showing that the state remains unchanged if the function is constant, while it can be in an arbitrary superposition if it is balanced. Finally, encoding the Hadamard matrix  and applying the cylindrical lens and performing the filtering leaves the y-coordinate in the state
\begin{equation}
\ket{v}=
\begin{cases}
\pm \ket{y_0}, & \text{Constant} \\
 \frac{1}{\sqrt{N}} \sum_{{j\neq0} }  u_j\ket{y_j} & \text{Balanced},
\end{cases}
\end{equation}
illustrated graphically in  Fig.~\ref{fig:Joza_concept} (b), showing the expected output intensities mapped onto the vertical rows of  the output field. For the constant case we have that the algorithm returns the first element of the basis , $\ket{v_0} \equiv \ket{y_0}$.   In the N=4 binary example, this is similar to obtaining the $\ket{v_0} \equiv \ket{00}$  state. The balanced case returns an arbitrary superposition of the logical basis states with coefficients given by $u_j = \frac{1}{\sqrt{N}} \sum_{m } (-1)^{m \cdot j + F(j)}$ where $m \cdot j$ is the binary inner product using the binary representation of $m$ and $j$.  For example, for a balanced function that maps as $\{00, 11\}\rightarrow 1 $ and $\{01, 10\} \rightarrow 0 $, corresponds to a unitary
\begin{equation}
    U_F = \begin{pmatrix}
       - 1 & 0 & 0 & 0 \\
         0 & 1 & 0 & 0 \\
          0 & 0& 1& 0 \\
          0 & 0& 0& -1 
        
    \end{pmatrix},
\end{equation}
represented in the logical basis.  The output for this example is $-\ket{v_3} \equiv - \ket{11}$. In general, the output state is one of the logical basis states that exclude, $\ket{v_0} \equiv \ket{00}$ as shown in  Fig.~\ref{fig:Joza_concept} (b).  
In our setup, we execute the protocol by encoding $\ket{u_0}$ on the first SLM and then subsequently encode $\hat{M}  = \hat{H}_N \hat{U}$ onto the second SLM.
\begin{figure*}[t!]
		\centering		\includegraphics[width=1\linewidth]{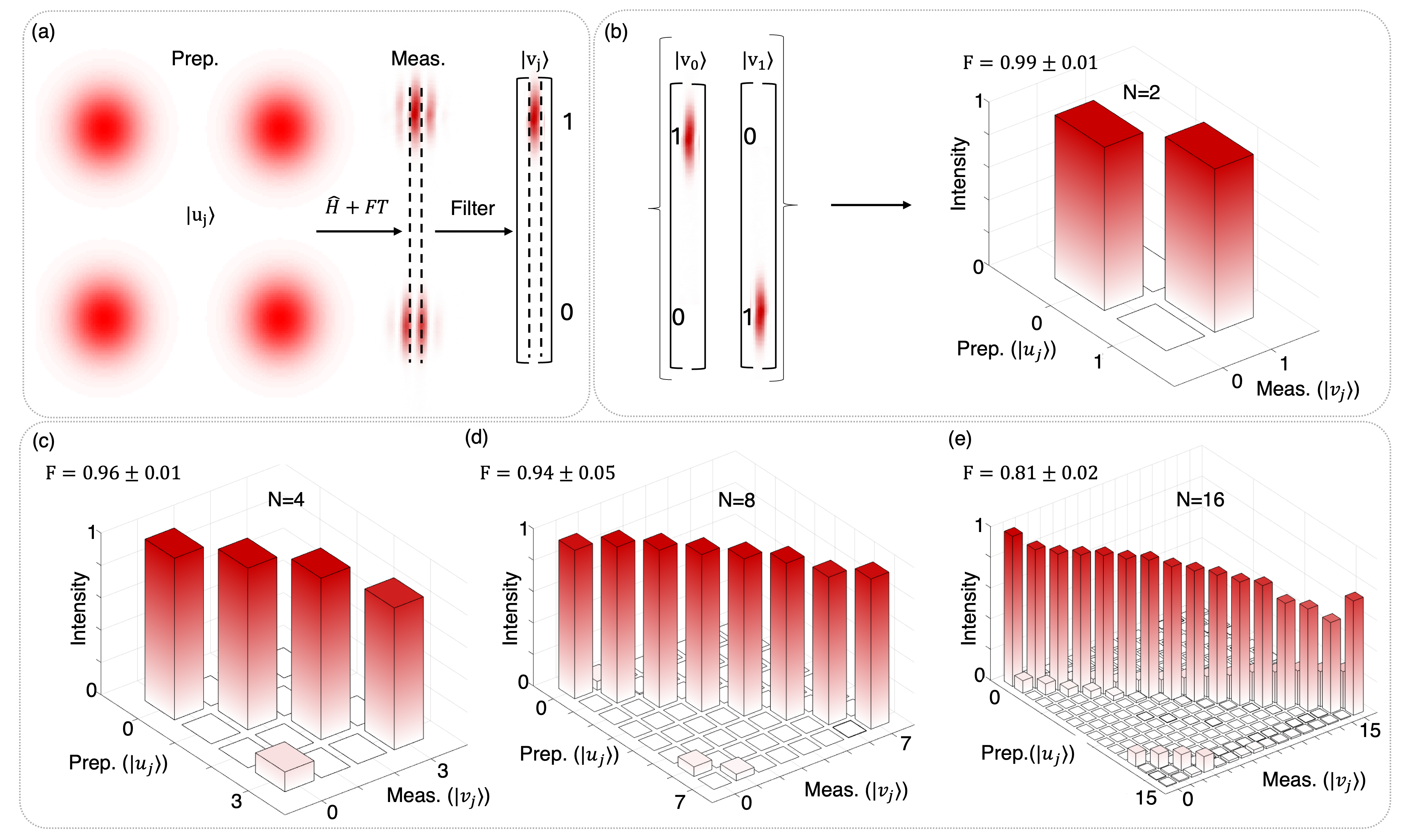}
  \caption{Setup Characterisation: Intensity images collected by the CCD camera when encoding a Hadamard matrix on the second SLM and the vector built from Hadamard columns on the first SLM. In (a), the extraction of the central lobe containing our solution vector from the multiplication of a 2x2 Hadamard matrix with the first column is highlighted. (b) We then stack these to form a diagonal matrix consisting of the light fringes as our elements. This diagonal matrix is then used to form a crosstalk matrix, from which the fidelity of the computation is computed. (c-e) Extension of this process to higher dimensional matrices, up to 16D.
  }
  \label{fig:CC}
	\end{figure*}
\section{Experiment}
 \begin{figure*}[ht!]
		\centering		\includegraphics[width=1\linewidth]{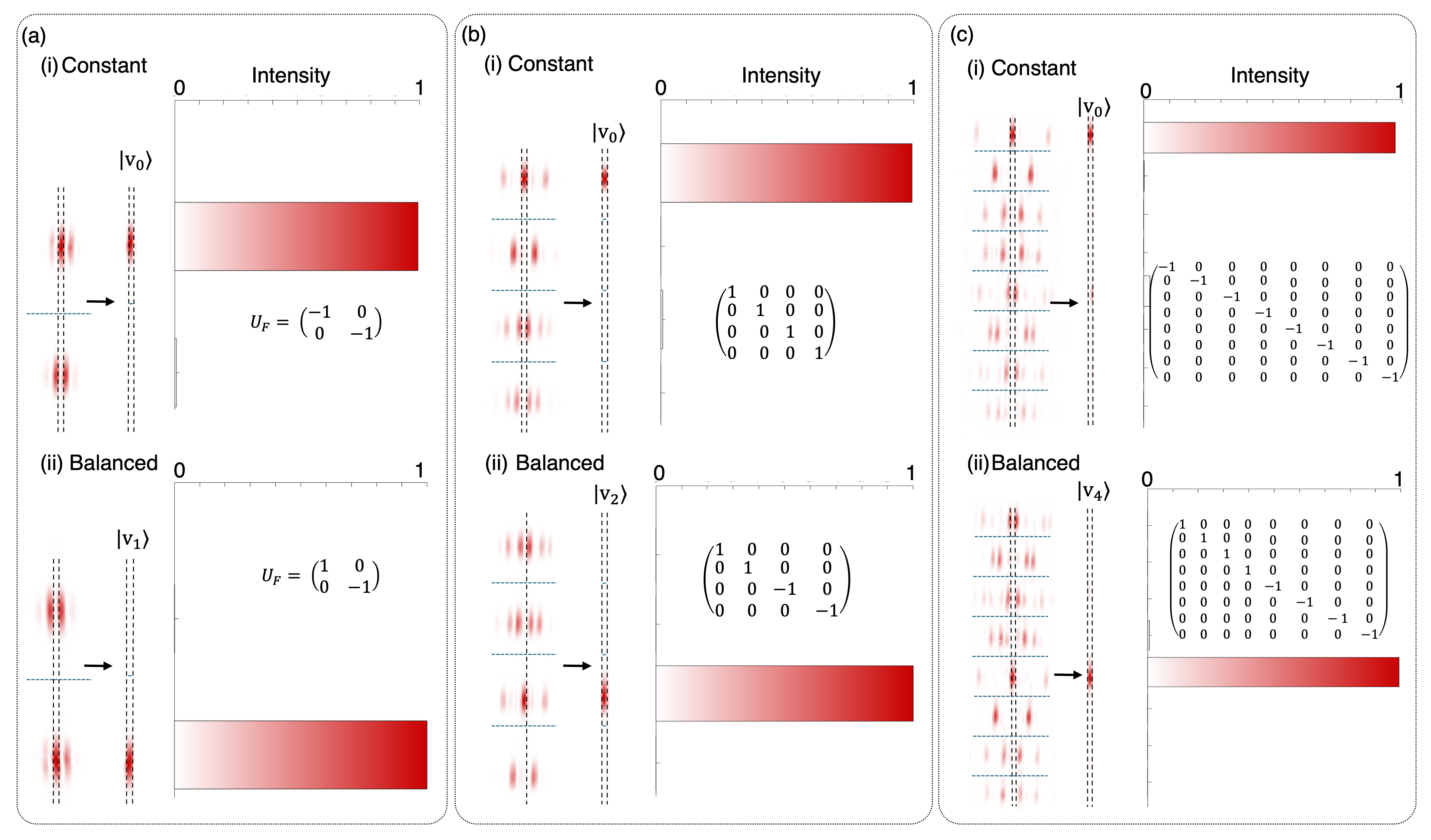}
  \caption{Deutsch-Jozsa algorithm results: 
 Algorithm results for the Deutsch-Jozsa algorithm implementation. Panels (a), (b), and (c) show the intensity distributions for single-qubit, two-qubit, and three-qubit systems, respectively, with the form of the oracle diagonal matrix $U_{F}$. In each panel: (1) Unfiltered images representing the initial light distribution, and (2) Filtered intensity images corresponding to the central fringe at \(k=0\). Subpanels (i) correspond to the constant function, and subpanels (ii) correspond to the balanced function. The bar graphs depict the normalized intensity against the row position, allowing the determination of the function type. The presence or absence of light in the first vector element's spatial position indicates whether the function is balanced or constant.}
  \label{fig:CF}
  \end{figure*}
In Fig.  \ref{fig:setup}  we illustrate the setup used for our demonstration. To ensure a uniform field intensity across the transverse plane, we overfilled the first spatial light modulator (SLM$_1$) by magnifying a laser beam from a Helium Neon (HeNe)  source with a wavelength of 633 nm, using the telescope. On $\text{SLM}_1$, the input vector for the lattice of Gaussian modes was encoded as an array of displaced Gaussian beams, producing the superposition state, $\ket{\Psi}\equiv\ket{u_0}$ from Eq.~(\ref{eq:lattice0}). To alter the coefficients of the lattice, each row of the matrix  (U)  mapping the vector was encoded using a digital hologram that encodes amplitudes and phases \cite{arrizon2007pixelated} onto each of the Gaussian modes. Because the SLM encodes the vector state  as a matrix, $U$, given any state $\ket{u}$, the elements were copied into its rows following the procedure outlined in the theory.

The matrix generated on $\text{SLM}_1$ was then transferred to $\text{SLM}_2$ using an imaging system consisting of lenses  $L_1$ and $L_2$. An aperture (AP) was placed between the two lenses to filter the modulated first order diffraction pattern from $\text{SLM}_1$.  On $\text{SLM}_2$, we applied an additional phase encoding to the lattice/matrix elements, thereby completing the optical representation of the matrix-vector multiplication process. Following the encoding on $\text{SLM}_2$, the beam was passed through a cylindrical lens (CL) to undergo a one-dimensional Fourier transform. This operation focused the outcome of the matrix-vector multiplication into the central fringe (shown as an inset in Fig. \ref{fig:setup}), specifically at the zeroth order Fourier component in coordinate space. To extract this outcome accurately, we employed digital filtering, enabling precise capture of the resultant vector as shown in the inset of Fig. \ref{fig:setup}.

\vspace{0.25in}

\section{Results}

\subsection{Matrix-Multiplication}
 To characterise our system, we exploit the orthogonality of the Hadamard basis. In order to do this, we prepared the Hadamard basis states ${\ket{u_j}}$ on $\text{SLM}_1$ and encoded the Hadamard transform on 
 $\text{SLM}_2$. Thereafter, we propagated the field through the cylindrical lens and collected the resulting interference pattern on a CCD camera. The multiplication between the Hadamard basis states  and the Hadamard gate enable us to clearly distinguish between different columns of the output vector state. Accordingly, we obtain orthogonal states ($\ket{v_j}$) with distinct positions marked by bright lobes. We show a measured example of this for $N=2$ dimensions  in Fig. \ref{fig:CC} (a) where  the first basis state, $\ket{u_0}$, is encoded on the lattice and measured using the Hadamard gate. The outcome is a measured interference pattern, where the resulting filtered zero order produces a bright lobe in the first entry - for this  specific example, the result corresponds to the state $\ket{v_0}$. Summing and normalising the intensity for each region of the output vector, corresponded to extracting the output vector elements.
 
 We repeated the process for all the basis states for a given dimension. In Fig.  \ref{fig:CC} (b)  we show the measured lobes for all the states in $N=2$, confirming that the Hadamard basis states ($\ket{u_0}$ and $\ket{u_1}$) are mapped on to the basis states  $\ket{v_0}$ and $\ket{v_1}$. From the measured lobes, we extracted the intensity from each component and produce a crosstalk matrix (C.M). For $N=2$. The cross talk matrix resembles an identity matrix showing that the measure basis modes are distinguishable and therefore confirms orthogonality between the measured vectors. To quantify this, we measured a fidelity of F = $0.99 \pm 0.01$, showing that the Hadamard matrix can map the basis elements {$\ket{u_j}$} onto the basis states {$\ket{v_j}$}. This was calculated by normalising each row, and computing the average over the diagonal components.

 Figures.~\ref{fig:CC} (c)-(e) show the extension of this process to higher-dimensional matrices. For dimensions $N=4$, $N=8$ and $N=16$ we obtain average fidelities of $0.96 \pm 0.01$, $0.94 \pm 0.05$ and $0.81 \pm 0.02$ multiplication, respectively. However, as the dimensions increase, we note that the fidelity decreases. We attribute this to the diffraction of Gaussian beams. Increasing the dimension or number of Gaussian modes for the same physical size of the matrix means decreasing the beam waist \(w_0\) of each mode. As the beam waists get smaller, they expand at a faster rate during propagation, according to $w(z) = w_0 \sqrt{1 + \left(\frac{\lambda z}{\pi w_0^2}\right)^2}$. Given that the cylindrical lens focuses light in one direction while neglecting the other, we expect the unfocused direction to continue expanding, more rapidly as \(w_0\) gets smaller, resulting in overlapping vector elements which are hard to distinguish. This diffraction effect impacts the fidelity of higher-dimensional computations.
 
\subsection{Deutsch-Jozsa Algorithm demonstration}

Using our scheme, we demonstrate  the Deutsch-Jozsa algorithm for a function that encodes information using $N=2$, $N=4$ and $N=8$ basis elements, shown in  Fig. \ref{fig:CF} (a)-(c), respectively. We initialized our system in the superposition state $\ket{u_0}$  by digitally displaying a hologram encoding the uniform lattice of Gaussian modes on $\text{SLM}_1$. The action of the oracle phase transformation, and the subsequent Hadamard gate are encoded as a compound unitary transformation $H \hat{U}_F $ on the second ($\text{SLM}_2$), the light beam is then sent through a cylindrical lens which acts as a summing operator. The resulting interference pattern that was measured with the CCD camera and filtered zero order vector components are shown as insets in each figure.  

From these images, we then plot the cross-section of the normalized intensity against the vector element number. In this case, we can distinguish whether the function is balanced or constant by interrogating the presence or absence of light in the spatial position of the vector element.  Using our basis vectors from the characterisation, this means a constant function produces the vector $\ket{v_0}$ and if the function is balanced  $\ket{v_{j \neq 0}}$.  The instance of the oracle operator ($\hat{U}_F$) used for each type of function is shown in each figure; this is  shown as an inset in the intensity plots representing the measured components.

For each case, we accurately determined whether the function encoded in the unitary operation is constant or balanced. The average fidelity of these measurements exceeded $90\%$ across all cases. Specifically, for N=2, 4, and 8, we achieved fidelities of $0.99 \pm 0.01$, $0.97 \pm 0.01$, and $0.93 \pm 0.05$, respectively.

\section{Discussion and Conclusion}

In this study, we utilized optical matrix-vector multiplication to emulate quantum computing by treating the $x$-$y$ components of a coherent field as a tensor product space, \(\mathcal{H}_N \otimes \mathcal{H}_N\), derived from individual coordinates. The encoding basis was defined within this Hilbert space using a lattice of Gaussian modes, where the positions of the centers in the $x$-$y$ coordinates established localized states. Vector states were encoded along the $x$-coordinate, while both coordinates were employed to encode the matrix operator. The output vector was measured along the $y$-coordinate after filtering out the zero-order component of the output interference pattern. We demonstrated the use of Hadamard basis elements and the Hadamard gate, achieving encoding and decoding fidelities of high as $95\%$ for $N=8$ dimensions and above $80\%$ for $N=16$.

Further, our implementation of the Deutsch-Jozsa algorithm using optical vector matrix multipliers leverages the representation of quantum states as vectors and quantum operators as matrices that can be imprinted on our lattice basis. Optically, we prepare multiple identical Gaussian modes in parallel, with  each mode  representing an element of the states that are inputs to the function $F(k)$.   The high-dimensional nature of our encoding scheme enables the oracle function in our demonstration to be encoded as a diagonal operator.

However, we note  spatial resources required to perform computations using structured light in higher dimensions are fundamentally limited by the diffraction limit. The minimum separation between distinguishable spatial modes is constrained by the wavelength of light and the numerical aperture of the optical system, leading to crosstalk between modes when attempting to scale up to higher dimensions. In our current setup,  we have explored up to  16 distinguishable spatial modes, equivalent to a 4-qubit system. J. Spall’s vector-matrix demonstration reached up to $\approx$ 50 dimensions of encoding - in principle, every pixel of the SLM can be utalised as an encoding channel but would require much effort . Moreover, as the number of spatial modes increases, the complexity of addressing and manipulating these modes grows significantly. In quantum systems, the number of operations required for a computation scales polynomially with the number of qubits, whereas in classical systems, the need to individually resolve and address each spatial mode results in a linear scaling with the number of available modes. This ultimately limits the scalability of classical systems in performing quantum-like operations.

Additionally the preparation of a Gaussian lattices and their unitary transformation matrices  on an SLM typically involves NxN operations to generate the necessary hologram for the spot array, which can be computationally expensive and inefficient when scaling to higher dimensions. An alternative approach is the use of fan-out operations to generate the lattice of Gaussian modes \cite{kirk1991design, zhou1992modified}. Techniques such as cylindrical lenses \cite{tamir2009high} or diffractive optical elements \cite{dammann1971high} can directly produce the desired lattice by splitting or reshaping the input beam, reducing the need for pixel-by-pixel modulation on the SLM and significantly improving efficiency. Recent advances in trained diffractive optical elements provide another potential solution by allowing computationally designed optical masks to perform the necessary optical transformations \cite{goel2024inverse}. This approach leverages inverse design techniques to optimize optical circuits, embedding high-dimensional transformations in complex media without requiring precise control over individual elements. Such masks streamline the process by reducing the need for real-time computation and re-calibration, ensuring scalability and maintaining the accuracy of the transformations.

We acknowledge the proposal by Perez-Garcia et al. \cite{perez2016quantum}, which employs classical light following Dragoman's method. However, this approach necessitates multiple optical elements to prepare the superposition state. In contrast, our method efficiently prepares the state in a single step by projecting light onto a spatial light modulator (SLM) in a reprogrammable manner -  the digital encoding allows for dynamic control and can therefore be adapted for any operation. Therefore, we anticipate that any transformation (operator/gate) can be encoded, including  X-gate, Z-gate, ect., because the method allows any matrix operation to be encoded into the lattice.

\begin{acknowledgments}
The authors acknowledge Prof. Andrew Forbes for his fruitful discussions and valuable advice. The authors also acknowledge the South African Quantum Initiative (SAQuti), the National Research Foundation (South Africa), and Optica for funding.

\end{acknowledgments}

\section*{Author Declaration}
The authors have no conflicts to disclose.

\section*{Data Availability Statement}
The data that support the findings of
this study are available from the
corresponding author upon reasonable
request

\section{References}
\nocite{*}


\begin{thebibliography}{55}%
\makeatletter
\providecommand \@ifxundefined [1]{%
 \@ifx{#1\undefined}
}%
\providecommand \@ifnum [1]{%
 \ifnum #1\expandafter \@firstoftwo
 \else \expandafter \@secondoftwo
 \fi
}%
\providecommand \@ifx [1]{%
 \ifx #1\expandafter \@firstoftwo
 \else \expandafter \@secondoftwo
 \fi
}%
\providecommand \natexlab [1]{#1}%
\providecommand \enquote  [1]{``#1''}%
\providecommand \bibnamefont  [1]{#1}%
\providecommand \bibfnamefont [1]{#1}%
\providecommand \citenamefont [1]{#1}%
\providecommand \href@noop [0]{\@secondoftwo}%
\providecommand \href [0]{\begingroup \@sanitize@url \@href}%
\providecommand \@href[1]{\@@startlink{#1}\@@href}%
\providecommand \@@href[1]{\endgroup#1\@@endlink}%
\providecommand \@sanitize@url [0]{\catcode `\\12\catcode `\$12\catcode `\&12\catcode `\#12\catcode `\^12\catcode `\_12\catcode `\%12\relax}%
\providecommand \@@startlink[1]{}%
\providecommand \@@endlink[0]{}%
\providecommand \url  [0]{\begingroup\@sanitize@url \@url }%
\providecommand \@url [1]{\endgroup\@href {#1}{\urlprefix }}%
\providecommand \urlprefix  [0]{URL }%
\providecommand \Eprint [0]{\href }%
\providecommand \doibase [0]{http://dx.doi.org/}%
\providecommand \selectlanguage [0]{\@gobble}%
\providecommand \bibinfo  [0]{\@secondoftwo}%
\providecommand \bibfield  [0]{\@secondoftwo}%
\providecommand \translation [1]{[#1]}%
\providecommand \BibitemOpen [0]{}%
\providecommand \bibitemStop [0]{}%
\providecommand \bibitemNoStop [0]{.\EOS\space}%
\providecommand \EOS [0]{\spacefactor3000\relax}%
\providecommand \BibitemShut  [1]{\csname bibitem#1\endcsname}%
\let\auto@bib@innerbib\@empty
\bibitem [{\citenamefont {Cameron}\ \emph {et~al.}(2024)\citenamefont {Cameron}, \citenamefont {Courme}, \citenamefont {Verni{\`e}re}, \citenamefont {Pandya}, \citenamefont {Faccio},\ and\ \citenamefont {Defienne}}]{cameron2024adaptive}%
  \BibitemOpen
  \bibfield  {author} {\bibinfo {author} {\bibfnamefont {P.}~\bibnamefont {Cameron}}, \bibinfo {author} {\bibfnamefont {B.}~\bibnamefont {Courme}}, \bibinfo {author} {\bibfnamefont {C.}~\bibnamefont {Verni{\`e}re}}, \bibinfo {author} {\bibfnamefont {R.}~\bibnamefont {Pandya}}, \bibinfo {author} {\bibfnamefont {D.}~\bibnamefont {Faccio}}, \ and\ \bibinfo {author} {\bibfnamefont {H.}~\bibnamefont {Defienne}},\ }\bibfield  {title} {\enquote {\bibinfo {title} {Adaptive optical imaging with entangled photons},}\ }\href@noop {} {\bibfield  {journal} {\bibinfo  {journal} {Science}\ }\textbf {\bibinfo {volume} {383}},\ \bibinfo {pages} {1142--1148} (\bibinfo {year} {2024})}\BibitemShut {NoStop}%
\bibitem [{\citenamefont {Moodley}\ and\ \citenamefont {Forbes}(2024)}]{moodley2024advances}%
  \BibitemOpen
  \bibfield  {author} {\bibinfo {author} {\bibfnamefont {C.}~\bibnamefont {Moodley}}\ and\ \bibinfo {author} {\bibfnamefont {A.}~\bibnamefont {Forbes}},\ }\bibfield  {title} {\enquote {\bibinfo {title} {Advances in quantum imaging with machine intelligence},}\ }\href@noop {} {\bibfield  {journal} {\bibinfo  {journal} {Laser \& Photonics Reviews}\ ,\ \bibinfo {pages} {2300939}} (\bibinfo {year} {2024})}\BibitemShut {NoStop}%
\bibitem [{\citenamefont {Geng}(2011)}]{geng2011structured}%
  \BibitemOpen
  \bibfield  {author} {\bibinfo {author} {\bibfnamefont {J.}~\bibnamefont {Geng}},\ }\bibfield  {title} {\enquote {\bibinfo {title} {Structured-light 3d surface imaging: a tutorial},}\ }\href@noop {} {\bibfield  {journal} {\bibinfo  {journal} {Advances in optics and photonics}\ }\textbf {\bibinfo {volume} {3}},\ \bibinfo {pages} {128--160} (\bibinfo {year} {2011})}\BibitemShut {NoStop}%
\bibitem [{\citenamefont {Erkmen}\ and\ \citenamefont {Shapiro}(2010)}]{erkmen2010ghost}%
  \BibitemOpen
  \bibfield  {author} {\bibinfo {author} {\bibfnamefont {B.~I.}\ \bibnamefont {Erkmen}}\ and\ \bibinfo {author} {\bibfnamefont {J.~H.}\ \bibnamefont {Shapiro}},\ }\bibfield  {title} {\enquote {\bibinfo {title} {Ghost imaging: from quantum to classical to computational},}\ }\href@noop {} {\bibfield  {journal} {\bibinfo  {journal} {Advances in Optics and Photonics}\ }\textbf {\bibinfo {volume} {2}},\ \bibinfo {pages} {405--450} (\bibinfo {year} {2010})}\BibitemShut {NoStop}%
\bibitem [{\citenamefont {Ferri}\ \emph {et~al.}(2010)\citenamefont {Ferri}, \citenamefont {Magatti}, \citenamefont {Lugiato},\ and\ \citenamefont {Gatti}}]{ferri2010differential}%
  \BibitemOpen
  \bibfield  {author} {\bibinfo {author} {\bibfnamefont {F.}~\bibnamefont {Ferri}}, \bibinfo {author} {\bibfnamefont {D.}~\bibnamefont {Magatti}}, \bibinfo {author} {\bibfnamefont {L.}~\bibnamefont {Lugiato}}, \ and\ \bibinfo {author} {\bibfnamefont {A.}~\bibnamefont {Gatti}},\ }\bibfield  {title} {\enquote {\bibinfo {title} {Differential ghost imaging},}\ }\href@noop {} {\bibfield  {journal} {\bibinfo  {journal} {Physical review letters}\ }\textbf {\bibinfo {volume} {104}},\ \bibinfo {pages} {253603} (\bibinfo {year} {2010})}\BibitemShut {NoStop}%
\bibitem [{\citenamefont {Zia}\ \emph {et~al.}(2023)\citenamefont {Zia}, \citenamefont {Dehghan}, \citenamefont {D’Errico}, \citenamefont {Sciarrino},\ and\ \citenamefont {Karimi}}]{zia2023interferometric}%
  \BibitemOpen
  \bibfield  {author} {\bibinfo {author} {\bibfnamefont {D.}~\bibnamefont {Zia}}, \bibinfo {author} {\bibfnamefont {N.}~\bibnamefont {Dehghan}}, \bibinfo {author} {\bibfnamefont {A.}~\bibnamefont {D’Errico}}, \bibinfo {author} {\bibfnamefont {F.}~\bibnamefont {Sciarrino}}, \ and\ \bibinfo {author} {\bibfnamefont {E.}~\bibnamefont {Karimi}},\ }\bibfield  {title} {\enquote {\bibinfo {title} {Interferometric imaging of amplitude and phase of spatial biphoton states},}\ }\href@noop {} {\bibfield  {journal} {\bibinfo  {journal} {Nature Photonics}\ }\textbf {\bibinfo {volume} {17}},\ \bibinfo {pages} {1009--1016} (\bibinfo {year} {2023})}\BibitemShut {NoStop}%
\bibitem [{\citenamefont {Kim}\ \emph {et~al.}(2022)\citenamefont {Kim}, \citenamefont {Kim}, \citenamefont {Yun}, \citenamefont {Moon}, \citenamefont {Kim}, \citenamefont {Kim}, \citenamefont {Park}, \citenamefont {Badloe}, \citenamefont {Kim},\ and\ \citenamefont {Rho}}]{kim2022metasurface}%
  \BibitemOpen
  \bibfield  {author} {\bibinfo {author} {\bibfnamefont {G.}~\bibnamefont {Kim}}, \bibinfo {author} {\bibfnamefont {Y.}~\bibnamefont {Kim}}, \bibinfo {author} {\bibfnamefont {J.}~\bibnamefont {Yun}}, \bibinfo {author} {\bibfnamefont {S.-W.}\ \bibnamefont {Moon}}, \bibinfo {author} {\bibfnamefont {S.}~\bibnamefont {Kim}}, \bibinfo {author} {\bibfnamefont {J.}~\bibnamefont {Kim}}, \bibinfo {author} {\bibfnamefont {J.}~\bibnamefont {Park}}, \bibinfo {author} {\bibfnamefont {T.}~\bibnamefont {Badloe}}, \bibinfo {author} {\bibfnamefont {I.}~\bibnamefont {Kim}}, \ and\ \bibinfo {author} {\bibfnamefont {J.}~\bibnamefont {Rho}},\ }\bibfield  {title} {\enquote {\bibinfo {title} {Metasurface-driven full-space structured light for three-dimensional imaging},}\ }\href@noop {} {\bibfield  {journal} {\bibinfo  {journal} {Nature Communications}\ }\textbf {\bibinfo {volume} {13}},\ \bibinfo {pages} {5920} (\bibinfo {year} {2022})}\BibitemShut {NoStop}%
\bibitem [{\citenamefont {Mirhosseini}\ \emph {et~al.}(2015)\citenamefont {Mirhosseini}, \citenamefont {Maga{\~n}a-Loaiza}, \citenamefont {O’Sullivan}, \citenamefont {Rodenburg}, \citenamefont {Malik}, \citenamefont {Lavery}, \citenamefont {Padgett}, \citenamefont {Gauthier},\ and\ \citenamefont {Boyd}}]{mirhosseini2015high}%
  \BibitemOpen
  \bibfield  {author} {\bibinfo {author} {\bibfnamefont {M.}~\bibnamefont {Mirhosseini}}, \bibinfo {author} {\bibfnamefont {O.~S.}\ \bibnamefont {Maga{\~n}a-Loaiza}}, \bibinfo {author} {\bibfnamefont {M.~N.}\ \bibnamefont {O’Sullivan}}, \bibinfo {author} {\bibfnamefont {B.}~\bibnamefont {Rodenburg}}, \bibinfo {author} {\bibfnamefont {M.}~\bibnamefont {Malik}}, \bibinfo {author} {\bibfnamefont {M.~P.}\ \bibnamefont {Lavery}}, \bibinfo {author} {\bibfnamefont {M.~J.}\ \bibnamefont {Padgett}}, \bibinfo {author} {\bibfnamefont {D.~J.}\ \bibnamefont {Gauthier}}, \ and\ \bibinfo {author} {\bibfnamefont {R.~W.}\ \bibnamefont {Boyd}},\ }\bibfield  {title} {\enquote {\bibinfo {title} {High-dimensional quantum cryptography with twisted light},}\ }\href@noop {} {\bibfield  {journal} {\bibinfo  {journal} {New Journal of Physics}\ }\textbf {\bibinfo {volume} {17}},\ \bibinfo {pages} {033033} (\bibinfo {year} {2015})}\BibitemShut {NoStop}%
\bibitem [{\citenamefont {Wang}\ \emph {et~al.}(2012)\citenamefont {Wang}, \citenamefont {Yang}, \citenamefont {Fazal}, \citenamefont {Ahmed}, \citenamefont {Yan}, \citenamefont {Huang}, \citenamefont {Ren}, \citenamefont {Yue}, \citenamefont {Dolinar}, \citenamefont {Tur} \emph {et~al.}}]{wang2012terabit}%
  \BibitemOpen
  \bibfield  {author} {\bibinfo {author} {\bibfnamefont {J.}~\bibnamefont {Wang}}, \bibinfo {author} {\bibfnamefont {J.-Y.}\ \bibnamefont {Yang}}, \bibinfo {author} {\bibfnamefont {I.~M.}\ \bibnamefont {Fazal}}, \bibinfo {author} {\bibfnamefont {N.}~\bibnamefont {Ahmed}}, \bibinfo {author} {\bibfnamefont {Y.}~\bibnamefont {Yan}}, \bibinfo {author} {\bibfnamefont {H.}~\bibnamefont {Huang}}, \bibinfo {author} {\bibfnamefont {Y.}~\bibnamefont {Ren}}, \bibinfo {author} {\bibfnamefont {Y.}~\bibnamefont {Yue}}, \bibinfo {author} {\bibfnamefont {S.}~\bibnamefont {Dolinar}}, \bibinfo {author} {\bibfnamefont {M.}~\bibnamefont {Tur}},  \emph {et~al.},\ }\bibfield  {title} {\enquote {\bibinfo {title} {Terabit free-space data transmission employing orbital angular momentum multiplexing},}\ }\href@noop {} {\bibfield  {journal} {\bibinfo  {journal} {Nature photonics}\ }\textbf {\bibinfo {volume} {6}},\ \bibinfo {pages} {488--496} (\bibinfo {year} {2012})}\BibitemShut {NoStop}%
\bibitem [{\citenamefont {Krenn}\ \emph {et~al.}(2014)\citenamefont {Krenn}, \citenamefont {Fickler}, \citenamefont {Fink}, \citenamefont {Handsteiner}, \citenamefont {Malik}, \citenamefont {Scheidl}, \citenamefont {Ursin},\ and\ \citenamefont {Zeilinger}}]{krenn2014communication}%
  \BibitemOpen
  \bibfield  {author} {\bibinfo {author} {\bibfnamefont {M.}~\bibnamefont {Krenn}}, \bibinfo {author} {\bibfnamefont {R.}~\bibnamefont {Fickler}}, \bibinfo {author} {\bibfnamefont {M.}~\bibnamefont {Fink}}, \bibinfo {author} {\bibfnamefont {J.}~\bibnamefont {Handsteiner}}, \bibinfo {author} {\bibfnamefont {M.}~\bibnamefont {Malik}}, \bibinfo {author} {\bibfnamefont {T.}~\bibnamefont {Scheidl}}, \bibinfo {author} {\bibfnamefont {R.}~\bibnamefont {Ursin}}, \ and\ \bibinfo {author} {\bibfnamefont {A.}~\bibnamefont {Zeilinger}},\ }\bibfield  {title} {\enquote {\bibinfo {title} {Communication with spatially modulated light through turbulent air across vienna},}\ }\href@noop {} {\bibfield  {journal} {\bibinfo  {journal} {New Journal of Physics}\ }\textbf {\bibinfo {volume} {16}},\ \bibinfo {pages} {113028} (\bibinfo {year} {2014})}\BibitemShut {NoStop}%
\bibitem [{\citenamefont {Wang}\ \emph {et~al.}(2022)\citenamefont {Wang}, \citenamefont {Liu}, \citenamefont {Li}, \citenamefont {Zhao}, \citenamefont {Du},\ and\ \citenamefont {Zhu}}]{wang2022orbital}%
  \BibitemOpen
  \bibfield  {author} {\bibinfo {author} {\bibfnamefont {J.}~\bibnamefont {Wang}}, \bibinfo {author} {\bibfnamefont {J.}~\bibnamefont {Liu}}, \bibinfo {author} {\bibfnamefont {S.}~\bibnamefont {Li}}, \bibinfo {author} {\bibfnamefont {Y.}~\bibnamefont {Zhao}}, \bibinfo {author} {\bibfnamefont {J.}~\bibnamefont {Du}}, \ and\ \bibinfo {author} {\bibfnamefont {L.}~\bibnamefont {Zhu}},\ }\bibfield  {title} {\enquote {\bibinfo {title} {Orbital angular momentum and beyond in free-space optical communications},}\ }\href@noop {} {\bibfield  {journal} {\bibinfo  {journal} {Nanophotonics}\ }\textbf {\bibinfo {volume} {11}},\ \bibinfo {pages} {645--680} (\bibinfo {year} {2022})}\BibitemShut {NoStop}%
\bibitem [{\citenamefont {Forbes}\ \emph {et~al.}(2024)\citenamefont {Forbes}, \citenamefont {Youssef}, \citenamefont {Singh}, \citenamefont {Nape},\ and\ \citenamefont {Ung}}]{forbes2024quantum}%
  \BibitemOpen
  \bibfield  {author} {\bibinfo {author} {\bibfnamefont {A.}~\bibnamefont {Forbes}}, \bibinfo {author} {\bibfnamefont {M.}~\bibnamefont {Youssef}}, \bibinfo {author} {\bibfnamefont {S.}~\bibnamefont {Singh}}, \bibinfo {author} {\bibfnamefont {I.}~\bibnamefont {Nape}}, \ and\ \bibinfo {author} {\bibfnamefont {B.}~\bibnamefont {Ung}},\ }\bibfield  {title} {\enquote {\bibinfo {title} {Quantum cryptography with structured photons},}\ }\href@noop {} {\bibfield  {journal} {\bibinfo  {journal} {Applied Physics Letters}\ }\textbf {\bibinfo {volume} {124}} (\bibinfo {year} {2024})}\BibitemShut {NoStop}%
\bibitem [{\citenamefont {Lavery}\ \emph {et~al.}(2017)\citenamefont {Lavery}, \citenamefont {Peuntinger}, \citenamefont {G{\"u}nthner}, \citenamefont {Banzer}, \citenamefont {Elser}, \citenamefont {Boyd}, \citenamefont {Padgett}, \citenamefont {Marquardt},\ and\ \citenamefont {Leuchs}}]{lavery2017free}%
  \BibitemOpen
  \bibfield  {author} {\bibinfo {author} {\bibfnamefont {M.~P.}\ \bibnamefont {Lavery}}, \bibinfo {author} {\bibfnamefont {C.}~\bibnamefont {Peuntinger}}, \bibinfo {author} {\bibfnamefont {K.}~\bibnamefont {G{\"u}nthner}}, \bibinfo {author} {\bibfnamefont {P.}~\bibnamefont {Banzer}}, \bibinfo {author} {\bibfnamefont {D.}~\bibnamefont {Elser}}, \bibinfo {author} {\bibfnamefont {R.~W.}\ \bibnamefont {Boyd}}, \bibinfo {author} {\bibfnamefont {M.~J.}\ \bibnamefont {Padgett}}, \bibinfo {author} {\bibfnamefont {C.}~\bibnamefont {Marquardt}}, \ and\ \bibinfo {author} {\bibfnamefont {G.}~\bibnamefont {Leuchs}},\ }\bibfield  {title} {\enquote {\bibinfo {title} {Free-space propagation of high-dimensional structured optical fields in an urban environment},}\ }\href@noop {} {\bibfield  {journal} {\bibinfo  {journal} {Science Advances}\ }\textbf {\bibinfo {volume} {3}},\ \bibinfo {pages} {e1700552} (\bibinfo {year} {2017})}\BibitemShut {NoStop}%
\bibitem [{\citenamefont {Perez-Garcia}\ \emph {et~al.}(2018)\citenamefont {Perez-Garcia}, \citenamefont {Hernandez-Aranda}, \citenamefont {Forbes},\ and\ \citenamefont {Konrad}}]{perez2018first}%
  \BibitemOpen
  \bibfield  {author} {\bibinfo {author} {\bibfnamefont {B.}~\bibnamefont {Perez-Garcia}}, \bibinfo {author} {\bibfnamefont {R.~I.}\ \bibnamefont {Hernandez-Aranda}}, \bibinfo {author} {\bibfnamefont {A.}~\bibnamefont {Forbes}}, \ and\ \bibinfo {author} {\bibfnamefont {T.}~\bibnamefont {Konrad}},\ }\bibfield  {title} {\enquote {\bibinfo {title} {The first iteration of grover's algorithm using classical light with orbital angular momentum},}\ }\href@noop {} {\bibfield  {journal} {\bibinfo  {journal} {Journal of Modern Optics}\ }\textbf {\bibinfo {volume} {65}},\ \bibinfo {pages} {1942--1948} (\bibinfo {year} {2018})}\BibitemShut {NoStop}%
\bibitem [{\citenamefont {Peruzzo}\ \emph {et~al.}(2014)\citenamefont {Peruzzo}, \citenamefont {McClean}, \citenamefont {Shadbolt}, \citenamefont {Yung}, \citenamefont {Zhou}, \citenamefont {Love}, \citenamefont {Aspuru-Guzik},\ and\ \citenamefont {O’brien}}]{peruzzo2014variational}%
  \BibitemOpen
  \bibfield  {author} {\bibinfo {author} {\bibfnamefont {A.}~\bibnamefont {Peruzzo}}, \bibinfo {author} {\bibfnamefont {J.}~\bibnamefont {McClean}}, \bibinfo {author} {\bibfnamefont {P.}~\bibnamefont {Shadbolt}}, \bibinfo {author} {\bibfnamefont {M.-H.}\ \bibnamefont {Yung}}, \bibinfo {author} {\bibfnamefont {X.-Q.}\ \bibnamefont {Zhou}}, \bibinfo {author} {\bibfnamefont {P.~J.}\ \bibnamefont {Love}}, \bibinfo {author} {\bibfnamefont {A.}~\bibnamefont {Aspuru-Guzik}}, \ and\ \bibinfo {author} {\bibfnamefont {J.~L.}\ \bibnamefont {O’brien}},\ }\bibfield  {title} {\enquote {\bibinfo {title} {A variational eigenvalue solver on a photonic quantum processor},}\ }\href@noop {} {\bibfield  {journal} {\bibinfo  {journal} {Nature communications}\ }\textbf {\bibinfo {volume} {5}},\ \bibinfo {pages} {4213} (\bibinfo {year} {2014})}\BibitemShut {NoStop}%
\bibitem [{\citenamefont {Konno}\ \emph {et~al.}(2024)\citenamefont {Konno}, \citenamefont {Asavanant}, \citenamefont {Hanamura}, \citenamefont {Nagayoshi}, \citenamefont {Fukui}, \citenamefont {Sakaguchi}, \citenamefont {Ide}, \citenamefont {China}, \citenamefont {Yabuno}, \citenamefont {Miki} \emph {et~al.}}]{konno2024logical}%
  \BibitemOpen
  \bibfield  {author} {\bibinfo {author} {\bibfnamefont {S.}~\bibnamefont {Konno}}, \bibinfo {author} {\bibfnamefont {W.}~\bibnamefont {Asavanant}}, \bibinfo {author} {\bibfnamefont {F.}~\bibnamefont {Hanamura}}, \bibinfo {author} {\bibfnamefont {H.}~\bibnamefont {Nagayoshi}}, \bibinfo {author} {\bibfnamefont {K.}~\bibnamefont {Fukui}}, \bibinfo {author} {\bibfnamefont {A.}~\bibnamefont {Sakaguchi}}, \bibinfo {author} {\bibfnamefont {R.}~\bibnamefont {Ide}}, \bibinfo {author} {\bibfnamefont {F.}~\bibnamefont {China}}, \bibinfo {author} {\bibfnamefont {M.}~\bibnamefont {Yabuno}}, \bibinfo {author} {\bibfnamefont {S.}~\bibnamefont {Miki}},  \emph {et~al.},\ }\bibfield  {title} {\enquote {\bibinfo {title} {Logical states for fault-tolerant quantum computation with propagating light},}\ }\href@noop {} {\bibfield  {journal} {\bibinfo  {journal} {Science}\ }\textbf {\bibinfo {volume} {383}},\ \bibinfo {pages} {289--293} (\bibinfo {year} {2024})}\BibitemShut {NoStop}%
\bibitem [{\citenamefont {McMahon}(2023)}]{mcmahon2023physics}%
  \BibitemOpen
  \bibfield  {author} {\bibinfo {author} {\bibfnamefont {P.~L.}\ \bibnamefont {McMahon}},\ }\bibfield  {title} {\enquote {\bibinfo {title} {The physics of optical computing},}\ }\href@noop {} {\bibfield  {journal} {\bibinfo  {journal} {Nature Reviews Physics}\ }\textbf {\bibinfo {volume} {5}},\ \bibinfo {pages} {717--734} (\bibinfo {year} {2023})}\BibitemShut {NoStop}%
\bibitem [{\citenamefont {Nape}\ \emph {et~al.}(2023)\citenamefont {Nape}, \citenamefont {Sephton}, \citenamefont {Ornelas}, \citenamefont {Moodley},\ and\ \citenamefont {Forbes}}]{nape2023quantum}%
  \BibitemOpen
  \bibfield  {author} {\bibinfo {author} {\bibfnamefont {I.}~\bibnamefont {Nape}}, \bibinfo {author} {\bibfnamefont {B.}~\bibnamefont {Sephton}}, \bibinfo {author} {\bibfnamefont {P.}~\bibnamefont {Ornelas}}, \bibinfo {author} {\bibfnamefont {C.}~\bibnamefont {Moodley}}, \ and\ \bibinfo {author} {\bibfnamefont {A.}~\bibnamefont {Forbes}},\ }\bibfield  {title} {\enquote {\bibinfo {title} {Quantum structured light in high dimensions},}\ }\href@noop {} {\bibfield  {journal} {\bibinfo  {journal} {APL Photonics}\ }\textbf {\bibinfo {volume} {8}} (\bibinfo {year} {2023})}\BibitemShut {NoStop}%
\bibitem [{\citenamefont {Forbes}, \citenamefont {de~Oliveira},\ and\ \citenamefont {Dennis}(2021)}]{forbes2021structured}%
  \BibitemOpen
  \bibfield  {author} {\bibinfo {author} {\bibfnamefont {A.}~\bibnamefont {Forbes}}, \bibinfo {author} {\bibfnamefont {M.}~\bibnamefont {de~Oliveira}}, \ and\ \bibinfo {author} {\bibfnamefont {M.~R.}\ \bibnamefont {Dennis}},\ }\bibfield  {title} {\enquote {\bibinfo {title} {Structured light},}\ }\href@noop {} {\bibfield  {journal} {\bibinfo  {journal} {Nature Photonics}\ }\textbf {\bibinfo {volume} {15}},\ \bibinfo {pages} {253--262} (\bibinfo {year} {2021})}\BibitemShut {NoStop}%
\bibitem [{\citenamefont {Rubinsztein-Dunlop}\ \emph {et~al.}(2016)\citenamefont {Rubinsztein-Dunlop}, \citenamefont {Forbes}, \citenamefont {Berry}, \citenamefont {Dennis}, \citenamefont {Andrews}, \citenamefont {Mansuripur}, \citenamefont {Denz}, \citenamefont {Alpmann}, \citenamefont {Banzer}, \citenamefont {Bauer} \emph {et~al.}}]{rubinsztein2016roadmap}%
  \BibitemOpen
  \bibfield  {author} {\bibinfo {author} {\bibfnamefont {H.}~\bibnamefont {Rubinsztein-Dunlop}}, \bibinfo {author} {\bibfnamefont {A.}~\bibnamefont {Forbes}}, \bibinfo {author} {\bibfnamefont {M.~V.}\ \bibnamefont {Berry}}, \bibinfo {author} {\bibfnamefont {M.~R.}\ \bibnamefont {Dennis}}, \bibinfo {author} {\bibfnamefont {D.~L.}\ \bibnamefont {Andrews}}, \bibinfo {author} {\bibfnamefont {M.}~\bibnamefont {Mansuripur}}, \bibinfo {author} {\bibfnamefont {C.}~\bibnamefont {Denz}}, \bibinfo {author} {\bibfnamefont {C.}~\bibnamefont {Alpmann}}, \bibinfo {author} {\bibfnamefont {P.}~\bibnamefont {Banzer}}, \bibinfo {author} {\bibfnamefont {T.}~\bibnamefont {Bauer}},  \emph {et~al.},\ }\bibfield  {title} {\enquote {\bibinfo {title} {Roadmap on structured light},}\ }\href@noop {} {\bibfield  {journal} {\bibinfo  {journal} {Journal of Optics}\ }\textbf {\bibinfo {volume} {19}},\ \bibinfo {pages} {013001} (\bibinfo {year} {2016})}\BibitemShut {NoStop}%
\bibitem [{\citenamefont {Piccardo}\ \emph {et~al.}(2021)\citenamefont {Piccardo}, \citenamefont {Ginis}, \citenamefont {Forbes}, \citenamefont {Mahler}, \citenamefont {Friesem}, \citenamefont {Davidson}, \citenamefont {Ren}, \citenamefont {Dorrah}, \citenamefont {Capasso}, \citenamefont {Dullo} \emph {et~al.}}]{piccardo2021roadmap}%
  \BibitemOpen
  \bibfield  {author} {\bibinfo {author} {\bibfnamefont {M.}~\bibnamefont {Piccardo}}, \bibinfo {author} {\bibfnamefont {V.}~\bibnamefont {Ginis}}, \bibinfo {author} {\bibfnamefont {A.}~\bibnamefont {Forbes}}, \bibinfo {author} {\bibfnamefont {S.}~\bibnamefont {Mahler}}, \bibinfo {author} {\bibfnamefont {A.~A.}\ \bibnamefont {Friesem}}, \bibinfo {author} {\bibfnamefont {N.}~\bibnamefont {Davidson}}, \bibinfo {author} {\bibfnamefont {H.}~\bibnamefont {Ren}}, \bibinfo {author} {\bibfnamefont {A.~H.}\ \bibnamefont {Dorrah}}, \bibinfo {author} {\bibfnamefont {F.}~\bibnamefont {Capasso}}, \bibinfo {author} {\bibfnamefont {F.~T.}\ \bibnamefont {Dullo}},  \emph {et~al.},\ }\bibfield  {title} {\enquote {\bibinfo {title} {Roadmap on multimode light shaping},}\ }\href@noop {} {\bibfield  {journal} {\bibinfo  {journal} {Journal of Optics}\ }\textbf {\bibinfo {volume} {24}},\ \bibinfo {pages} {013001} (\bibinfo {year} {2021})}\BibitemShut {NoStop}%
\bibitem [{\citenamefont {Bliokh}\ \emph {et~al.}(2023)\citenamefont {Bliokh}, \citenamefont {Karimi}, \citenamefont {Padgett}, \citenamefont {Alonso}, \citenamefont {Dennis}, \citenamefont {Dudley}, \citenamefont {Forbes}, \citenamefont {Zahedpour}, \citenamefont {Hancock}, \citenamefont {Milchberg} \emph {et~al.}}]{bliokh2023roadmap}%
  \BibitemOpen
  \bibfield  {author} {\bibinfo {author} {\bibfnamefont {K.~Y.}\ \bibnamefont {Bliokh}}, \bibinfo {author} {\bibfnamefont {E.}~\bibnamefont {Karimi}}, \bibinfo {author} {\bibfnamefont {M.~J.}\ \bibnamefont {Padgett}}, \bibinfo {author} {\bibfnamefont {M.~A.}\ \bibnamefont {Alonso}}, \bibinfo {author} {\bibfnamefont {M.~R.}\ \bibnamefont {Dennis}}, \bibinfo {author} {\bibfnamefont {A.}~\bibnamefont {Dudley}}, \bibinfo {author} {\bibfnamefont {A.}~\bibnamefont {Forbes}}, \bibinfo {author} {\bibfnamefont {S.}~\bibnamefont {Zahedpour}}, \bibinfo {author} {\bibfnamefont {S.~W.}\ \bibnamefont {Hancock}}, \bibinfo {author} {\bibfnamefont {H.~M.}\ \bibnamefont {Milchberg}},  \emph {et~al.},\ }\bibfield  {title} {\enquote {\bibinfo {title} {Roadmap on structured waves},}\ }\href@noop {} {\bibfield  {journal} {\bibinfo  {journal} {Journal of Optics}\ }\textbf {\bibinfo {volume} {25}},\ \bibinfo {pages} {103001} (\bibinfo {year} {2023})}\BibitemShut {NoStop}%
\bibitem [{\citenamefont {Boykin}\ \emph {et~al.}(2000)\citenamefont {Boykin}, \citenamefont {Mor}, \citenamefont {Pulver}, \citenamefont {Roychowdhury},\ and\ \citenamefont {Vatan}}]{boykin2000new}%
  \BibitemOpen
  \bibfield  {author} {\bibinfo {author} {\bibfnamefont {P.~O.}\ \bibnamefont {Boykin}}, \bibinfo {author} {\bibfnamefont {T.}~\bibnamefont {Mor}}, \bibinfo {author} {\bibfnamefont {M.}~\bibnamefont {Pulver}}, \bibinfo {author} {\bibfnamefont {V.}~\bibnamefont {Roychowdhury}}, \ and\ \bibinfo {author} {\bibfnamefont {F.}~\bibnamefont {Vatan}},\ }\bibfield  {title} {\enquote {\bibinfo {title} {A new universal and fault-tolerant quantum basis},}\ }\href@noop {} {\bibfield  {journal} {\bibinfo  {journal} {Information Processing Letters}\ }\textbf {\bibinfo {volume} {75}},\ \bibinfo {pages} {101--107} (\bibinfo {year} {2000})}\BibitemShut {NoStop}%
\bibitem [{\citenamefont {Kitaev}, \citenamefont {Shen},\ and\ \citenamefont {Vyalyi}(2002)}]{kitaev2002classical}%
  \BibitemOpen
  \bibfield  {author} {\bibinfo {author} {\bibfnamefont {A.~Y.}\ \bibnamefont {Kitaev}}, \bibinfo {author} {\bibfnamefont {A.}~\bibnamefont {Shen}}, \ and\ \bibinfo {author} {\bibfnamefont {M.~N.}\ \bibnamefont {Vyalyi}},\ }\href@noop {} {\emph {\bibinfo {title} {Classical and quantum computation}}},\ \bibinfo {number} {47}\ (\bibinfo  {publisher} {American Mathematical Soc.},\ \bibinfo {year} {2002})\BibitemShut {NoStop}%
\bibitem [{\citenamefont {Wang}\ \emph {et~al.}(2020)\citenamefont {Wang}, \citenamefont {Hu}, \citenamefont {Sanders},\ and\ \citenamefont {Kais}}]{wang2020qudits}%
  \BibitemOpen
  \bibfield  {author} {\bibinfo {author} {\bibfnamefont {Y.}~\bibnamefont {Wang}}, \bibinfo {author} {\bibfnamefont {Z.}~\bibnamefont {Hu}}, \bibinfo {author} {\bibfnamefont {B.~C.}\ \bibnamefont {Sanders}}, \ and\ \bibinfo {author} {\bibfnamefont {S.}~\bibnamefont {Kais}},\ }\bibfield  {title} {\enquote {\bibinfo {title} {Qudits and high-dimensional quantum computing},}\ }\href@noop {} {\bibfield  {journal} {\bibinfo  {journal} {Frontiers in Physics}\ }\textbf {\bibinfo {volume} {8}},\ \bibinfo {pages} {589504} (\bibinfo {year} {2020})}\BibitemShut {NoStop}%
\bibitem [{\citenamefont {Ecker}\ \emph {et~al.}(2019)\citenamefont {Ecker}, \citenamefont {Bouchard}, \citenamefont {Bulla}, \citenamefont {Brandt}, \citenamefont {Kohout}, \citenamefont {Steinlechner}, \citenamefont {Fickler}, \citenamefont {Malik}, \citenamefont {Guryanova}, \citenamefont {Ursin} \emph {et~al.}}]{ecker2019overcoming}%
  \BibitemOpen
  \bibfield  {author} {\bibinfo {author} {\bibfnamefont {S.}~\bibnamefont {Ecker}}, \bibinfo {author} {\bibfnamefont {F.}~\bibnamefont {Bouchard}}, \bibinfo {author} {\bibfnamefont {L.}~\bibnamefont {Bulla}}, \bibinfo {author} {\bibfnamefont {F.}~\bibnamefont {Brandt}}, \bibinfo {author} {\bibfnamefont {O.}~\bibnamefont {Kohout}}, \bibinfo {author} {\bibfnamefont {F.}~\bibnamefont {Steinlechner}}, \bibinfo {author} {\bibfnamefont {R.}~\bibnamefont {Fickler}}, \bibinfo {author} {\bibfnamefont {M.}~\bibnamefont {Malik}}, \bibinfo {author} {\bibfnamefont {Y.}~\bibnamefont {Guryanova}}, \bibinfo {author} {\bibfnamefont {R.}~\bibnamefont {Ursin}},  \emph {et~al.},\ }\bibfield  {title} {\enquote {\bibinfo {title} {Overcoming noise in entanglement distribution},}\ }\href@noop {} {\bibfield  {journal} {\bibinfo  {journal} {Physical Review X}\ }\textbf {\bibinfo {volume} {9}},\ \bibinfo {pages} {041042} (\bibinfo {year} {2019})}\BibitemShut {NoStop}%
\bibitem [{\citenamefont {Spall}\ \emph {et~al.}(2020)\citenamefont {Spall}, \citenamefont {Guo}, \citenamefont {Barrett},\ and\ \citenamefont {Lvovsky}}]{spall2020fully}%
  \BibitemOpen
  \bibfield  {author} {\bibinfo {author} {\bibfnamefont {J.}~\bibnamefont {Spall}}, \bibinfo {author} {\bibfnamefont {X.}~\bibnamefont {Guo}}, \bibinfo {author} {\bibfnamefont {T.~D.}\ \bibnamefont {Barrett}}, \ and\ \bibinfo {author} {\bibfnamefont {A.}~\bibnamefont {Lvovsky}},\ }\bibfield  {title} {\enquote {\bibinfo {title} {Fully reconfigurable coherent optical vector--matrix multiplication},}\ }\href@noop {} {\bibfield  {journal} {\bibinfo  {journal} {Optics Letters}\ }\textbf {\bibinfo {volume} {45}},\ \bibinfo {pages} {5752--5755} (\bibinfo {year} {2020})}\BibitemShut {NoStop}%
\bibitem [{\citenamefont {Garcia-Escartin}\ and\ \citenamefont {Chamorro-Posada}(2012)}]{garcia2012quantum}%
  \BibitemOpen
  \bibfield  {author} {\bibinfo {author} {\bibfnamefont {J.~C.}\ \bibnamefont {Garcia-Escartin}}\ and\ \bibinfo {author} {\bibfnamefont {P.}~\bibnamefont {Chamorro-Posada}},\ }\bibfield  {title} {\enquote {\bibinfo {title} {Quantum computer networks with the orbital angular momentum of light},}\ }\href@noop {} {\bibfield  {journal} {\bibinfo  {journal} {Physical Review A}\ }\textbf {\bibinfo {volume} {86}},\ \bibinfo {pages} {032334} (\bibinfo {year} {2012})}\BibitemShut {NoStop}%
\bibitem [{\citenamefont {O'brien}(2007)}]{o2007optical}%
  \BibitemOpen
  \bibfield  {author} {\bibinfo {author} {\bibfnamefont {J.~L.}\ \bibnamefont {O'brien}},\ }\bibfield  {title} {\enquote {\bibinfo {title} {Optical quantum computing},}\ }\href@noop {} {\bibfield  {journal} {\bibinfo  {journal} {Science}\ }\textbf {\bibinfo {volume} {318}},\ \bibinfo {pages} {1567--1570} (\bibinfo {year} {2007})}\BibitemShut {NoStop}%
\bibitem [{\citenamefont {Gao}\ \emph {et~al.}(2019)\citenamefont {Gao}, \citenamefont {Krenn}, \citenamefont {Kysela},\ and\ \citenamefont {Zeilinger}}]{gao2019arbitrary}%
  \BibitemOpen
  \bibfield  {author} {\bibinfo {author} {\bibfnamefont {X.}~\bibnamefont {Gao}}, \bibinfo {author} {\bibfnamefont {M.}~\bibnamefont {Krenn}}, \bibinfo {author} {\bibfnamefont {J.}~\bibnamefont {Kysela}}, \ and\ \bibinfo {author} {\bibfnamefont {A.}~\bibnamefont {Zeilinger}},\ }\bibfield  {title} {\enquote {\bibinfo {title} {Arbitrary d-dimensional pauli x gates of a flying qudit},}\ }\href@noop {} {\bibfield  {journal} {\bibinfo  {journal} {Physical review A}\ }\textbf {\bibinfo {volume} {99}},\ \bibinfo {pages} {023825} (\bibinfo {year} {2019})}\BibitemShut {NoStop}%
\bibitem [{\citenamefont {Fontaine}\ \emph {et~al.}(2022)\citenamefont {Fontaine}, \citenamefont {Carpenter}, \citenamefont {Gross}, \citenamefont {Leon-Saval}, \citenamefont {Jung}, \citenamefont {Richardson},\ and\ \citenamefont {Amezcua-Correa}}]{fontaine2022photonic}%
  \BibitemOpen
  \bibfield  {author} {\bibinfo {author} {\bibfnamefont {N.~K.}\ \bibnamefont {Fontaine}}, \bibinfo {author} {\bibfnamefont {J.}~\bibnamefont {Carpenter}}, \bibinfo {author} {\bibfnamefont {S.}~\bibnamefont {Gross}}, \bibinfo {author} {\bibfnamefont {S.}~\bibnamefont {Leon-Saval}}, \bibinfo {author} {\bibfnamefont {Y.}~\bibnamefont {Jung}}, \bibinfo {author} {\bibfnamefont {D.~J.}\ \bibnamefont {Richardson}}, \ and\ \bibinfo {author} {\bibfnamefont {R.}~\bibnamefont {Amezcua-Correa}},\ }\bibfield  {title} {\enquote {\bibinfo {title} {Photonic lanterns, 3-d waveguides, multiplane light conversion, and other components that enable space-division multiplexing},}\ }\href@noop {} {\bibfield  {journal} {\bibinfo  {journal} {Proceedings of the IEEE}\ }\textbf {\bibinfo {volume} {110}},\ \bibinfo {pages} {1821--1834} (\bibinfo {year} {2022})}\BibitemShut {NoStop}%
\bibitem [{\citenamefont {Brandt}\ \emph {et~al.}(2020)\citenamefont {Brandt}, \citenamefont {Hiekkam{\"a}ki}, \citenamefont {Bouchard}, \citenamefont {Huber},\ and\ \citenamefont {Fickler}}]{brandt2020high}%
  \BibitemOpen
  \bibfield  {author} {\bibinfo {author} {\bibfnamefont {F.}~\bibnamefont {Brandt}}, \bibinfo {author} {\bibfnamefont {M.}~\bibnamefont {Hiekkam{\"a}ki}}, \bibinfo {author} {\bibfnamefont {F.}~\bibnamefont {Bouchard}}, \bibinfo {author} {\bibfnamefont {M.}~\bibnamefont {Huber}}, \ and\ \bibinfo {author} {\bibfnamefont {R.}~\bibnamefont {Fickler}},\ }\bibfield  {title} {\enquote {\bibinfo {title} {High-dimensional quantum gates using full-field spatial modes of photons},}\ }\href@noop {} {\bibfield  {journal} {\bibinfo  {journal} {Optica}\ }\textbf {\bibinfo {volume} {7}},\ \bibinfo {pages} {98--107} (\bibinfo {year} {2020})}\BibitemShut {NoStop}%
\bibitem [{\citenamefont {Wang}\ \emph {et~al.}(2024)\citenamefont {Wang}, \citenamefont {Liu}, \citenamefont {Lyu},\ and\ \citenamefont {Wang}}]{wang2024ultrahigh}%
  \BibitemOpen
  \bibfield  {author} {\bibinfo {author} {\bibfnamefont {Q.}~\bibnamefont {Wang}}, \bibinfo {author} {\bibfnamefont {J.}~\bibnamefont {Liu}}, \bibinfo {author} {\bibfnamefont {D.}~\bibnamefont {Lyu}}, \ and\ \bibinfo {author} {\bibfnamefont {J.}~\bibnamefont {Wang}},\ }\bibfield  {title} {\enquote {\bibinfo {title} {Ultrahigh-fidelity spatial mode quantum gates in high-dimensional space by diffractive deep neural networks},}\ }\href@noop {} {\bibfield  {journal} {\bibinfo  {journal} {Light: Science \& Applications}\ }\textbf {\bibinfo {volume} {13}},\ \bibinfo {pages} {10} (\bibinfo {year} {2024})}\BibitemShut {NoStop}%
\bibitem [{\citenamefont {Lib}\ and\ \citenamefont {Bromberg}(2022)}]{lib2022quantum}%
  \BibitemOpen
  \bibfield  {author} {\bibinfo {author} {\bibfnamefont {O.}~\bibnamefont {Lib}}\ and\ \bibinfo {author} {\bibfnamefont {Y.}~\bibnamefont {Bromberg}},\ }\bibfield  {title} {\enquote {\bibinfo {title} {Quantum light in complex media and its applications},}\ }\href@noop {} {\bibfield  {journal} {\bibinfo  {journal} {Nature Physics}\ }\textbf {\bibinfo {volume} {18}},\ \bibinfo {pages} {986--993} (\bibinfo {year} {2022})}\BibitemShut {NoStop}%
\bibitem [{\citenamefont {Goel}\ \emph {et~al.}(2024)\citenamefont {Goel}, \citenamefont {Leedumrongwatthanakun}, \citenamefont {Valencia}, \citenamefont {McCutcheon}, \citenamefont {Tavakoli}, \citenamefont {Conti}, \citenamefont {Pinkse},\ and\ \citenamefont {Malik}}]{goel2024inverse}%
  \BibitemOpen
  \bibfield  {author} {\bibinfo {author} {\bibfnamefont {S.}~\bibnamefont {Goel}}, \bibinfo {author} {\bibfnamefont {S.}~\bibnamefont {Leedumrongwatthanakun}}, \bibinfo {author} {\bibfnamefont {N.~H.}\ \bibnamefont {Valencia}}, \bibinfo {author} {\bibfnamefont {W.}~\bibnamefont {McCutcheon}}, \bibinfo {author} {\bibfnamefont {A.}~\bibnamefont {Tavakoli}}, \bibinfo {author} {\bibfnamefont {C.}~\bibnamefont {Conti}}, \bibinfo {author} {\bibfnamefont {P.~W.}\ \bibnamefont {Pinkse}}, \ and\ \bibinfo {author} {\bibfnamefont {M.}~\bibnamefont {Malik}},\ }\bibfield  {title} {\enquote {\bibinfo {title} {Inverse design of high-dimensional quantum optical circuits in a complex medium},}\ }\href@noop {} {\bibfield  {journal} {\bibinfo  {journal} {Nature Physics}\ ,\ \bibinfo {pages} {1--8}} (\bibinfo {year} {2024})}\BibitemShut {NoStop}%
\bibitem [{\citenamefont {Cerf}, \citenamefont {Adami},\ and\ \citenamefont {Kwiat}(1998)}]{cerf1998optical}%
  \BibitemOpen
  \bibfield  {author} {\bibinfo {author} {\bibfnamefont {N.~J.}\ \bibnamefont {Cerf}}, \bibinfo {author} {\bibfnamefont {C.}~\bibnamefont {Adami}}, \ and\ \bibinfo {author} {\bibfnamefont {P.~G.}\ \bibnamefont {Kwiat}},\ }\bibfield  {title} {\enquote {\bibinfo {title} {Optical simulation of quantum logic},}\ }\href@noop {} {\bibfield  {journal} {\bibinfo  {journal} {Physical Review A}\ }\textbf {\bibinfo {volume} {57}},\ \bibinfo {pages} {R1477} (\bibinfo {year} {1998})}\BibitemShut {NoStop}%
\bibitem [{\citenamefont {Spreeuw}(2001)}]{spreeuw2001classical}%
  \BibitemOpen
  \bibfield  {author} {\bibinfo {author} {\bibfnamefont {R.~J.}\ \bibnamefont {Spreeuw}},\ }\bibfield  {title} {\enquote {\bibinfo {title} {Classical wave-optics analogy of quantum-information processing},}\ }\href@noop {} {\bibfield  {journal} {\bibinfo  {journal} {Physical Review A}\ }\textbf {\bibinfo {volume} {63}},\ \bibinfo {pages} {062302} (\bibinfo {year} {2001})}\BibitemShut {NoStop}%
\bibitem [{\citenamefont {Londero}\ \emph {et~al.}(2004)\citenamefont {Londero}, \citenamefont {Dorrer}, \citenamefont {Anderson}, \citenamefont {Wallentowitz}, \citenamefont {Banaszek},\ and\ \citenamefont {Walmsley}}]{londero2004efficient}%
  \BibitemOpen
  \bibfield  {author} {\bibinfo {author} {\bibfnamefont {P.}~\bibnamefont {Londero}}, \bibinfo {author} {\bibfnamefont {C.}~\bibnamefont {Dorrer}}, \bibinfo {author} {\bibfnamefont {M.}~\bibnamefont {Anderson}}, \bibinfo {author} {\bibfnamefont {S.}~\bibnamefont {Wallentowitz}}, \bibinfo {author} {\bibfnamefont {K.}~\bibnamefont {Banaszek}}, \ and\ \bibinfo {author} {\bibfnamefont {I.}~\bibnamefont {Walmsley}},\ }\bibfield  {title} {\enquote {\bibinfo {title} {Efficient optical implementation of the bernstein-vazirani algorithm},}\ }\href@noop {} {\bibfield  {journal} {\bibinfo  {journal} {Physical Review A}\ }\textbf {\bibinfo {volume} {69}},\ \bibinfo {pages} {010302} (\bibinfo {year} {2004})}\BibitemShut {NoStop}%
\bibitem [{\citenamefont {Kaur}, \citenamefont {Narang}\ \emph {et~al.}(2007)\citenamefont {Kaur}, \citenamefont {Narang} \emph {et~al.}}]{kaur2007optical}%
  \BibitemOpen
  \bibfield  {author} {\bibinfo {author} {\bibfnamefont {G.}~\bibnamefont {Kaur}}, \bibinfo {author} {\bibfnamefont {G.}~\bibnamefont {Narang}},  \emph {et~al.},\ }\bibfield  {title} {\enquote {\bibinfo {title} {Optical implementations, oracle equivalence, and the bernstein-vazirani algorithm},}\ }\href@noop {} {\bibfield  {journal} {\bibinfo  {journal} {JOSA B}\ }\textbf {\bibinfo {volume} {24}},\ \bibinfo {pages} {221--225} (\bibinfo {year} {2007})}\BibitemShut {NoStop}%
\bibitem [{\citenamefont {Perez-Garcia}\ \emph {et~al.}(2016)\citenamefont {Perez-Garcia}, \citenamefont {McLaren}, \citenamefont {Goyal}, \citenamefont {Hernandez-Aranda}, \citenamefont {Forbes},\ and\ \citenamefont {Konrad}}]{perez2016quantum}%
  \BibitemOpen
  \bibfield  {author} {\bibinfo {author} {\bibfnamefont {B.}~\bibnamefont {Perez-Garcia}}, \bibinfo {author} {\bibfnamefont {M.}~\bibnamefont {McLaren}}, \bibinfo {author} {\bibfnamefont {S.~K.}\ \bibnamefont {Goyal}}, \bibinfo {author} {\bibfnamefont {R.~I.}\ \bibnamefont {Hernandez-Aranda}}, \bibinfo {author} {\bibfnamefont {A.}~\bibnamefont {Forbes}}, \ and\ \bibinfo {author} {\bibfnamefont {T.}~\bibnamefont {Konrad}},\ }\bibfield  {title} {\enquote {\bibinfo {title} {Quantum computation with classical light: implementation of the deutsch--jozsa algorithm},}\ }\href@noop {} {\bibfield  {journal} {\bibinfo  {journal} {Physics Letters A}\ }\textbf {\bibinfo {volume} {380}},\ \bibinfo {pages} {1925--1931} (\bibinfo {year} {2016})}\BibitemShut {NoStop}%
\bibitem [{\citenamefont {Jozsa}(1997)}]{jozsa1997entanglement}%
  \BibitemOpen
  \bibfield  {author} {\bibinfo {author} {\bibfnamefont {R.}~\bibnamefont {Jozsa}},\ }\bibfield  {title} {\enquote {\bibinfo {title} {Entanglement and quantum computation},}\ }\href@noop {} {\bibfield  {journal} {\bibinfo  {journal} {arXiv preprint quant-ph/9707034}\ } (\bibinfo {year} {1997})}\BibitemShut {NoStop}%
\bibitem [{\citenamefont {Jozsa}\ and\ \citenamefont {Linden}(2003)}]{jozsa2003role}%
  \BibitemOpen
  \bibfield  {author} {\bibinfo {author} {\bibfnamefont {R.}~\bibnamefont {Jozsa}}\ and\ \bibinfo {author} {\bibfnamefont {N.}~\bibnamefont {Linden}},\ }\bibfield  {title} {\enquote {\bibinfo {title} {On the role of entanglement in quantum-computational speed-up},}\ }\href@noop {} {\bibfield  {journal} {\bibinfo  {journal} {Proceedings of the Royal Society of London. Series A: Mathematical, Physical and Engineering Sciences}\ }\textbf {\bibinfo {volume} {459}},\ \bibinfo {pages} {2011--2032} (\bibinfo {year} {2003})}\BibitemShut {NoStop}%
\bibitem [{\citenamefont {Wang}\ \emph {et~al.}(2017)\citenamefont {Wang}, \citenamefont {Poto{\v{c}}ek}, \citenamefont {Barnett},\ and\ \citenamefont {Feng}}]{wang2017programmable}%
  \BibitemOpen
  \bibfield  {author} {\bibinfo {author} {\bibfnamefont {Y.}~\bibnamefont {Wang}}, \bibinfo {author} {\bibfnamefont {V.}~\bibnamefont {Poto{\v{c}}ek}}, \bibinfo {author} {\bibfnamefont {S.~M.}\ \bibnamefont {Barnett}}, \ and\ \bibinfo {author} {\bibfnamefont {X.}~\bibnamefont {Feng}},\ }\bibfield  {title} {\enquote {\bibinfo {title} {Programmable holographic technique for implementing unitary and nonunitary transformations},}\ }\href@noop {} {\bibfield  {journal} {\bibinfo  {journal} {Physical Review A}\ }\textbf {\bibinfo {volume} {95}},\ \bibinfo {pages} {033827} (\bibinfo {year} {2017})}\BibitemShut {NoStop}%
\bibitem [{\citenamefont {Nitta}, \citenamefont {Matoba},\ and\ \citenamefont {Yoshimura}(2008)}]{nitta2008parallel}%
  \BibitemOpen
  \bibfield  {author} {\bibinfo {author} {\bibfnamefont {K.}~\bibnamefont {Nitta}}, \bibinfo {author} {\bibfnamefont {O.}~\bibnamefont {Matoba}}, \ and\ \bibinfo {author} {\bibfnamefont {T.}~\bibnamefont {Yoshimura}},\ }\bibfield  {title} {\enquote {\bibinfo {title} {Parallel processing for multiplication modulo by means of phase modulation},}\ }\href@noop {} {\bibfield  {journal} {\bibinfo  {journal} {Applied optics}\ }\textbf {\bibinfo {volume} {47}},\ \bibinfo {pages} {611--616} (\bibinfo {year} {2008})}\BibitemShut {NoStop}%
\bibitem [{\citenamefont {Tamura}\ and\ \citenamefont {Wyant}(1979)}]{tamura1979two}%
  \BibitemOpen
  \bibfield  {author} {\bibinfo {author} {\bibfnamefont {P.~N.}\ \bibnamefont {Tamura}}\ and\ \bibinfo {author} {\bibfnamefont {J.~C.}\ \bibnamefont {Wyant}},\ }\bibfield  {title} {\enquote {\bibinfo {title} {Two-dimensional matrix multiplication using coherent optical techniques},}\ }\href@noop {} {\bibfield  {journal} {\bibinfo  {journal} {Optical Engineering}\ }\textbf {\bibinfo {volume} {18}},\ \bibinfo {pages} {198--204} (\bibinfo {year} {1979})}\BibitemShut {NoStop}%
\bibitem [{\citenamefont {Horn}(1990)}]{horn1990hadamard}%
  \BibitemOpen
  \bibfield  {author} {\bibinfo {author} {\bibfnamefont {R.~A.}\ \bibnamefont {Horn}},\ }\bibfield  {title} {\enquote {\bibinfo {title} {The hadamard product},}\ }in\ \href@noop {} {\emph {\bibinfo {booktitle} {Proc. Symp. Appl. Math}}},\ Vol.~\bibinfo {volume} {40}\ (\bibinfo {year} {1990})\ pp.\ \bibinfo {pages} {87--169}\BibitemShut {NoStop}%
\bibitem [{\citenamefont {Shen}\ and\ \citenamefont {Rosales-Guzm{\'a}n}(2022)}]{shen2022nonseparable}%
  \BibitemOpen
  \bibfield  {author} {\bibinfo {author} {\bibfnamefont {Y.}~\bibnamefont {Shen}}\ and\ \bibinfo {author} {\bibfnamefont {C.}~\bibnamefont {Rosales-Guzm{\'a}n}},\ }\bibfield  {title} {\enquote {\bibinfo {title} {Nonseparable states of light: from quantum to classical},}\ }\href@noop {} {\bibfield  {journal} {\bibinfo  {journal} {Laser \& Photonics Reviews}\ }\textbf {\bibinfo {volume} {16}},\ \bibinfo {pages} {2100533} (\bibinfo {year} {2022})}\BibitemShut {NoStop}%
\bibitem [{\citenamefont {Aiello}\ \emph {et~al.}(2015)\citenamefont {Aiello}, \citenamefont {T{\"o}ppel}, \citenamefont {Marquardt}, \citenamefont {Giacobino},\ and\ \citenamefont {Leuchs}}]{aiello2015quantum}%
  \BibitemOpen
  \bibfield  {author} {\bibinfo {author} {\bibfnamefont {A.}~\bibnamefont {Aiello}}, \bibinfo {author} {\bibfnamefont {F.}~\bibnamefont {T{\"o}ppel}}, \bibinfo {author} {\bibfnamefont {C.}~\bibnamefont {Marquardt}}, \bibinfo {author} {\bibfnamefont {E.}~\bibnamefont {Giacobino}}, \ and\ \bibinfo {author} {\bibfnamefont {G.}~\bibnamefont {Leuchs}},\ }\bibfield  {title} {\enquote {\bibinfo {title} {Quantum- like nonseparable structures in optical beams},}\ }\href@noop {} {\bibfield  {journal} {\bibinfo  {journal} {New Journal of Physics}\ }\textbf {\bibinfo {volume} {17}},\ \bibinfo {pages} {043024} (\bibinfo {year} {2015})}\BibitemShut {NoStop}%
\bibitem [{\citenamefont {He}, \citenamefont {Shen},\ and\ \citenamefont {Forbes}(2022)}]{he2022towards}%
  \BibitemOpen
  \bibfield  {author} {\bibinfo {author} {\bibfnamefont {C.}~\bibnamefont {He}}, \bibinfo {author} {\bibfnamefont {Y.}~\bibnamefont {Shen}}, \ and\ \bibinfo {author} {\bibfnamefont {A.}~\bibnamefont {Forbes}},\ }\bibfield  {title} {\enquote {\bibinfo {title} {Towards higher-dimensional structured light},}\ }\href@noop {} {\bibfield  {journal} {\bibinfo  {journal} {Light: Science \& Applications}\ }\textbf {\bibinfo {volume} {11}},\ \bibinfo {pages} {205} (\bibinfo {year} {2022})}\BibitemShut {NoStop}%
\bibitem [{\citenamefont {Deutsch}\ and\ \citenamefont {Jozsa}(1992)}]{deutsch1992rapid}%
  \BibitemOpen
  \bibfield  {author} {\bibinfo {author} {\bibfnamefont {D.}~\bibnamefont {Deutsch}}\ and\ \bibinfo {author} {\bibfnamefont {R.}~\bibnamefont {Jozsa}},\ }\bibfield  {title} {\enquote {\bibinfo {title} {Rapid solution of problems by quantum computation},}\ }\href@noop {} {\bibfield  {journal} {\bibinfo  {journal} {Proceedings of the Royal Society of London. Series A: Mathematical and Physical Sciences}\ }\textbf {\bibinfo {volume} {439}},\ \bibinfo {pages} {553--558} (\bibinfo {year} {1992})}\BibitemShut {NoStop}%
\bibitem [{\citenamefont {Arriz{\'o}n}\ \emph {et~al.}(2007)\citenamefont {Arriz{\'o}n}, \citenamefont {Ruiz}, \citenamefont {Carrada},\ and\ \citenamefont {Gonz{\'a}lez}}]{arrizon2007pixelated}%
  \BibitemOpen
  \bibfield  {author} {\bibinfo {author} {\bibfnamefont {V.}~\bibnamefont {Arriz{\'o}n}}, \bibinfo {author} {\bibfnamefont {U.}~\bibnamefont {Ruiz}}, \bibinfo {author} {\bibfnamefont {R.}~\bibnamefont {Carrada}}, \ and\ \bibinfo {author} {\bibfnamefont {L.~A.}\ \bibnamefont {Gonz{\'a}lez}},\ }\bibfield  {title} {\enquote {\bibinfo {title} {Pixelated phase computer holograms for the accurate encoding of scalar complex fields},}\ }\href@noop {} {\bibfield  {journal} {\bibinfo  {journal} {JOSA A}\ }\textbf {\bibinfo {volume} {24}},\ \bibinfo {pages} {3500--3507} (\bibinfo {year} {2007})}\BibitemShut {NoStop}%
\bibitem [{\citenamefont {Kirk}\ \emph {et~al.}(1991)\citenamefont {Kirk}, \citenamefont {Imam}, \citenamefont {Bird},\ and\ \citenamefont {Hall}}]{kirk1991design}%
  \BibitemOpen
  \bibfield  {author} {\bibinfo {author} {\bibfnamefont {A.~G.}\ \bibnamefont {Kirk}}, \bibinfo {author} {\bibfnamefont {H.~T.}\ \bibnamefont {Imam}}, \bibinfo {author} {\bibfnamefont {K.}~\bibnamefont {Bird}}, \ and\ \bibinfo {author} {\bibfnamefont {T.~J.}\ \bibnamefont {Hall}},\ }\bibfield  {title} {\enquote {\bibinfo {title} {Design and fabrication of computer-generated holographic fan-out elements for a matrix/matrix interconnection scheme},}\ }in\ \href@noop {} {\emph {\bibinfo {booktitle} {Intl Colloquium on Diffractive Optical Elements}}},\ Vol.\ \bibinfo {volume} {1574}\ (\bibinfo {organization} {SPIE},\ \bibinfo {year} {1991})\ pp.\ \bibinfo {pages} {121--132}\BibitemShut {NoStop}%
\bibitem [{\citenamefont {Zhou}\ \emph {et~al.}(1992)\citenamefont {Zhou}, \citenamefont {Campbell}, \citenamefont {Yeh},\ and\ \citenamefont {Liu}}]{zhou1992modified}%
  \BibitemOpen
  \bibfield  {author} {\bibinfo {author} {\bibfnamefont {S.}~\bibnamefont {Zhou}}, \bibinfo {author} {\bibfnamefont {S.}~\bibnamefont {Campbell}}, \bibinfo {author} {\bibfnamefont {P.}~\bibnamefont {Yeh}}, \ and\ \bibinfo {author} {\bibfnamefont {H.-k.}\ \bibnamefont {Liu}},\ }\bibfield  {title} {\enquote {\bibinfo {title} {Modified-signed-digit optical computing by using fan-out elements},}\ }\href@noop {} {\bibfield  {journal} {\bibinfo  {journal} {Optics letters}\ }\textbf {\bibinfo {volume} {17}},\ \bibinfo {pages} {1697--1699} (\bibinfo {year} {1992})}\BibitemShut {NoStop}%
\bibitem [{\citenamefont {Tamir}\ \emph {et~al.}(2009)\citenamefont {Tamir}, \citenamefont {Shaked}, \citenamefont {Wilson},\ and\ \citenamefont {Dolev}}]{tamir2009high}%
  \BibitemOpen
  \bibfield  {author} {\bibinfo {author} {\bibfnamefont {D.~E.}\ \bibnamefont {Tamir}}, \bibinfo {author} {\bibfnamefont {N.~T.}\ \bibnamefont {Shaked}}, \bibinfo {author} {\bibfnamefont {P.~J.}\ \bibnamefont {Wilson}}, \ and\ \bibinfo {author} {\bibfnamefont {S.}~\bibnamefont {Dolev}},\ }\bibfield  {title} {\enquote {\bibinfo {title} {High-speed and low-power electro-optical dsp coprocessor},}\ }\href@noop {} {\bibfield  {journal} {\bibinfo  {journal} {JOSA A}\ }\textbf {\bibinfo {volume} {26}},\ \bibinfo {pages} {A11--A20} (\bibinfo {year} {2009})}\BibitemShut {NoStop}%
\bibitem [{\citenamefont {Dammann}\ and\ \citenamefont {G{\"o}rtler}(1971)}]{dammann1971high}%
  \BibitemOpen
  \bibfield  {author} {\bibinfo {author} {\bibfnamefont {H.}~\bibnamefont {Dammann}}\ and\ \bibinfo {author} {\bibfnamefont {K.}~\bibnamefont {G{\"o}rtler}},\ }\bibfield  {title} {\enquote {\bibinfo {title} {High-efficiency in-line multiple imaging by means of multiple phase holograms},}\ }\href@noop {} {\bibfield  {journal} {\bibinfo  {journal} {Optics communications}\ }\textbf {\bibinfo {volume} {3}},\ \bibinfo {pages} {312--315} (\bibinfo {year} {1971})}\BibitemShut {NoStop}%
\end{thebibliography}
\end{document}